\documentclass[prd,preprint,nofootinbib,12pt]{revtex4}
\usepackage{amsmath,amssymb}
\usepackage{enumitem}  
\usepackage[usenames, dvipsnames]{color} 
\usepackage{graphicx} 
\usepackage{comment}
\usepackage[normalem]{ulem}
\usepackage[utf8]{inputenc}
\UseRawInputEncoding 
\usepackage{url}
 \usepackage{xcolor}
 \usepackage{multirow}
 \usepackage{amssymb}
\usepackage{pifont}
\usepackage{amsthm}
\usepackage{hyperref}  

\usepackage{graphicx}
\usepackage{dcolumn}
\usepackage{bm}

\newcommand{\be}{\begin{equation}}
\newcommand{\ee}{\end{equation}}
\newcommand{\ba}{\begin{eqnarray}}
\newcommand{\ea}{\end{eqnarray}}

\newcommand{\beq}{\begin{equation}}
\newcommand{\eeq}{\end{equation}}
\newcommand{\beqa}{\begin{eqnarray}}
\newcommand{\eeqa}{\end{eqnarray}}

\newcommand{\dd}{{\rm{d}}}

\begin{document}


\title{Exact Kerr--Newman--(A)dS and other spacetimes in bumblebee gravity: employing a simple generating technique}%

\author{Hryhorii Ovcharenko}

\email{hryhorii.ovcharenko@matfyz.cuni.cz}

\affiliation{Institute of Theoretical Physics, Faculty of Mathematics and Physics,
Charles University, Prague, V Hole{\v s}ovi{\v c}k{\' a}ch 2, 180 00 Prague 8, Czech Republic}

\date{\today}

\begin{abstract}
In this work, we show that if the bumblebee field in the Einstein-bumblebee theory is given by its vacuum expectation value ($B_{\mu}=b_{\mu}$) and it is not dynamical ($\partial_{\mu}B_{\nu}-\partial_{\nu}B_{\mu}=0$), then these conditions \textit{uniquely} provide a generating technique, allowing us to construct exact solutions to bumblebee gravity from the vacuum solutions by adding a term $\sim b_{\mu}b_{\nu}$ to the metric tensor (thus proving the uniqueness of the method, presented in [Eur. Phys. J. C {\bf 82} (2022) 613]). Also, we show that the bumblebee field within this technique is proportional to the tangential vector of the (timelike or spacelike) geodesic curve in the background vacuum spacetime, and can be easily found knowing the solution to the Hamilton--Jacobi equation. Moreover, we prove that this technique can be extended to the case of any non-zero cosmological constant and the presence of the electromagnetic field. We apply this generating technique and obtain the bumblebee extension of the Kerr--Newman--Taub-NUT--(anti-)de Sitter spacetime. We show that this extension is not unique, as it depends on the exact geodesic curve one chooses to associate a bumblebee field with. Then, by considering various special cases of this generic solution, we demonstrate that the condition of the global reality of the bumblebee field limits the set of geodesics with which we can associate it.
\end{abstract}

 \maketitle

\newpage

\tableofcontents

\newpage

\section{Introduction}

General Relativity (GR), since its appearance, has been able to describe a lot of unusual, initially unexpected, spacetime geometries such as black holes, gravitational waves, etc. Even though a lot of these spacetimes were recently confirmed by observations, it still suffers from some conceptual issues. Namely, there is no unique description of quantum gravity that allows one to bypass the renormalizability issue of GR. However, among those that exist, they suggest the appearance of the effects of quantum gravity that can be observed even on the low energy scales, leading to the effective Lorentz symmetry breaking. It is worth mentioning that such effects were predicted by the loop quantum gravity (LQG) \cite{Gambini1999,Ellis2000} and the string theory \cite{Kosteleck1989,Kosteleck1989_2}.

It is obvious that the corresponding exact modifications of GR, emergent from existing quantum gravity models, are quite complicated on the classical level. Simplification of this analysis motivated the introduction of the effective field theory, which introduces an external bumblebee field $B^{\mu}$ that has some non-zero vacuum expectation value (VEV) $b^{\mu}$ that has to catch the Lorentz-violation effects. As VEV is the vector field, this directly introduces a Lorentz symmetry breaking on the level of a background spacetime, as there already exists a preferred direction in the spacetime, along which the bumblebee field is oriented. Such a theory was presented in \cite{Kostelecky2004}, where the authors presented a Standard Model Extension (SME) coupled to GR with the Lorentz symmetry breaking bumblebee field $B^{\mu}$. 

The most obvious application of the gravitational part of this model is to have a look at the black hole solutions and to study how they are modified by the bumblebee field. The first such Schwarzschild--like solution was obtained in \cite{Casana2018}, followed by the Schwarzschild-AdS spacetimes \cite{Maluf2021} and other exact solutions \cite{Liu2025,Li2025,Liu2025_2,Bailey2025,Xu2023,Marques2023,Filho2025_2}. These solutions (and the ones mentioned below) were used to investigate the influence of the bumblebee field on quasinormal modes and greybody factors \cite{Oliveira2021,Guo2024,Liu2023_qnm}, gravitational lensing \cite{Kuang2022,Filho2025_4}, thermodynamics \cite{Gomes2020}, and other aspects \cite{Pantig2025,Wang2022,Kanzi2019,Li2020,Khodadi2023,Shi2025,Filho_2025_1,Filho2025_3}. However, an attempt to find the exact Kerr-like spacetime in the bumblebee gravity initially failed \cite{Ding2020} as this solution does not satisfy the bumblebee field equation \cite{Maluf2022} and can only describe a slowly rotating limit. Nevertheless, this issue was bypassed in \cite{Poulis2022}. In this work the authors proposed a way of generating solutions to the bumblebee gravity in which--if one is able to find such a bumblebee field $B_{\mu}$, equal to its VEV value $B_{\mu}=b_{\mu}$ that the conditions $b^{\mu}b_{\mu}=b^2$ and $B_{\mu\nu}=2\nabla_{[\mu}B_{\nu]}=0$ are satisfied--then the metric $g_{\mu\nu}=\tilde{g}_{\mu\nu}+\dfrac{\xi}{1+\xi b^2}b_{\mu}b_{\nu}$, where $\tilde{g}_{\mu\nu}$ is the background vacuum spacetime, will be the solution to the field equations. This important result, in particular, was useful to generate the Kerr black hole within the bumblebee gravity. However, this work leaves several important questions that have not been answered yet. For example, it does not prove that the conditions $b^{\mu}b_{\mu}=b^2$ and $B_{\mu\nu}=2\nabla_{[\mu}B_{\nu]}=0$ generate the solution $g_{\mu\nu}=\tilde{g}_{\mu\nu}+\dfrac{\xi}{1+\xi b^2}b_{\mu}b_{\nu}$ \textit{uniquely}, and that there does not exist any other modification of the metric, satisfying the same conditions. Also, this work considers only the bumblebee field as a modification of the vacuum field equations without mentioning what happens when one adds a cosmological constant or some matter fields. In addition, in this work the authors did not present any algorithm that allows to find the bumblebee field $B_{\mu}=b_{\mu}$, satisfying the conditions $b^{\mu}b_{\nu}=b^2$ and $B_{\mu\nu}=2\nabla_{[\mu}B_{\nu]}=0$, leaving this as a separate mathematical issue. In principle, this can lead to complications if one deals with nontrivial spacetime. And the last thing is that this work did not investigate some of the solutions generated therein in great detail.

Our work aims to fulfill the gaps in the literature on these topics and to provide a proof that the conditions $b^{\mu}b_{\nu}=b^2$ and $B_{\mu\nu}=2\nabla_{[\mu}B_{\nu]}=0$ for $B_{\mu}=b_{\mu}$ \textit{uniquely} define the modification of the metric that satisfies the field equations. Also, we show that these conditions can be uniquely satisfied, knowing the solution to the Hamilton--Jacobi equation on the background metric, which allows one to easily generate the bumblebee field and the metric. We show that this generating technique also works in the presence of a non-zero cosmological constant and with the presence of the electromagnetic field. Additionally, we perform a physical analysis of the resulting solutions in more detail and show that there is a large class of novel solutions. However, the requirement for the bumblebee field to be \textit{globally real} strongly limits the possible range of generalizations.

The work is organized as follows. In Sec. \ref{Sec_der} we rederive the generating technique, found in \cite{Poulis2022}, show its uniqueness, and relate the task of finding the bumblebee field to the solution of the Hamilton--Jacobi equation on the ``seed'' spacetime. In Sec. \ref{Sec_generalizations} we show that this generating technique also works for the case with any cosmological constant and an electromagnetic field. In Sec. \ref{Sec_applic} we apply this generating technique to find the bumblebee version of the Kerr--Newman--Taub-NUT--(A)dS spacetime, investigate its various subcases, and discuss under which conditions the bumblebee field may be globally real. Concluding remarks are given in Sec.~\ref{Sec_concl}. 

\section{Proof of the uniqueness of the generating technique}\label{Sec_der}

\subsection{Review of the bumblebee gravity}

In this section, we briefly review the bumblebee gravity and present the generating technique that will be used further. The main feature of the bumblebee gravity is the presence of the vector field $B_{\mu}$ (called the bumblebee field) that acquires some non-zero vacuum expectation value (VEV) $b
_{\mu}$, which leads to Lorentz symmetry breaking because of the presence of some preferred directions of the spacetime. The action describing such a field, non-minimally coupled to gravity, was at first proposed in \cite{Kostelecky2004} and is given by
\begin{align}
    S=\int d^4x\sqrt{-g}\Big[\dfrac{1}{2\kappa}(R-2\Lambda)+\dfrac{\xi}{2\kappa}B^{\mu}B^{\nu}R_{\mu\nu}-\dfrac{1}{4}B_{\mu\nu}B^{\mu\nu}-V(B^{\mu}B_{\mu}-\epsilon b^2)\Big]+S_m,\label{bb_act}
\end{align}
where 
\begin{align}
    B_{\mu\nu}=\nabla_{\mu}B_{\nu}-\nabla_{\nu}B_{\mu},
\end{align}
$\Lambda$ is the cosmological constant, $\kappa=\dfrac{8\pi G}{c^4}$, and the parameter $\xi$ is the constant of non-minimal gravity and the bumblebee field couplings.

The potential $V$ governs the dynamics of the bumblebee field. As the main feature of this model is the presence of some non-zero vacuum expectation value, it is usually assumed that this potential has a minimum at $B^{\mu}B_{\nu}=\epsilon b^2$. Here $b$ is some constant, representing a strength of the bumblebee field, and $\epsilon=\pm 1$ represents whether this field is timelike or spacelike. $S_m$ is an action of the matter fields that we currently omit. 

By varying the action with respect to both the metric $g_{\mu\nu}$ and the bumblebee field $B_{\mu}$, one obtains the field equations for the metric
\begin{align}
    R_{\mu\nu}-\Lambda g_{\mu\nu}=&\kappa\Big[V'(2 B_{\mu}B_{\nu}-B_{\rho}B^{\rho} g_{\mu\nu})+B_{\mu}^{~\alpha}B_{\nu\alpha}+V g_{\mu\nu}-\dfrac{1}{4}B_{\alpha\beta}B^{\alpha\beta}g_{\mu\nu}\Big]\nonumber\\
    +&\xi\Big[\dfrac{1}{2}B^{\alpha}B^{\beta}R_{\alpha\beta}\,g_{\mu\nu}-B_{\mu}B^{\alpha}R_{\alpha \nu}-B_{\nu}B^{\alpha}R_{\alpha\mu}\Big]+\dfrac{\xi}{4}g_{\mu\nu}\nabla^2(B^{\alpha}B_{\alpha})\label{grav_eom}\\
    +&\dfrac{\xi}{2}\Big[\nabla_{\alpha}\nabla_{\mu}(B^{\alpha}B_{\nu})+\nabla_{\alpha}\nabla_{\nu}(B^{\alpha}B_{\mu})-\nabla^2(B_{\mu}B_{\nu})\Big],\nonumber
\end{align}
and for the bumblebee field
\begin{align}
    \nabla_{\mu}B^{\mu\nu}-2\Big(V'B^{\nu}-\dfrac{\xi}{2 \kappa}B_{\mu}R^{\mu\nu}\Big)=0.\label{bb_eom}
\end{align}

\subsection{Simplifying the field equations and the main part of the proof}

Let us start with the case of zero cosmological constant $\Lambda=0$ and without any matter fields. For the generating technique discussed in this work, we are interested in solutions when the bumblebee field $B_{\mu}$ remains frozen at its vacuum expectation value $b_{\mu}$. This ``freezing" condition also suggests that the bumblebee field takes the minimal value of the effective potential, so we also assume $V=0$ and $V'=0$. Not assuming these conditions may, in principle, lead to interesting results \cite{Bailey2025}, but we do not wish to study this here.

In addition, we assume another important property. Mainly, the works studying solutions in the bumblebee gravity investigated spherically symmetric spacetimes with the purely radial bumblebee field. These assumptions led to quite important condition $B_{\mu\nu}=\partial_{\mu}B_{\nu}-\partial_{\nu}B_{\mu}=0$ that allowed for the integration of the field equations. However, if one considers more general, not spherically symmetric, spacetimes, then the condition of purely radial $B_{\mu}$ breaks the property $B_{\mu\nu}=0$. Solutions in which this condition is not satisfied in principle exist \cite{Xu2023}, but this leads to the complication of the integration of the field equation that we would like to avoid.

For the generating technique, instead of some special choice of the direction of the bumblebee field $B_{\mu}$, we assume that the condition $B_{\mu\nu}=0$ holds.

Thus, here we assume several assumptions, namely
\begin{align}
    B_{\mu}=b_{\mu},~~~B_{\mu\nu}=\partial_{\mu}B_{\nu}-\partial_{\mu}B_{\mu}=0,\label{asum_1}
\end{align}
and
\begin{align}
    V=0,~~V'=0.\label{asum_2}
\end{align}

Thus, in the generating technique we discuss in this work, we have to solve equations (\ref{grav_eom})-(\ref{bb_eom}) with the assumptions (\ref{asum_1})-(\ref{asum_2}). These assumptions are quite strong as they directly simplify the field equations. First of all, we note that the field equation for the bumblebee field (\ref{bb_eom}) becomes 
\begin{align}
    b^{\mu}R_{\mu\nu}=0.\label{bb_eq1}
\end{align}

Substituting this equation and (\ref{asum_1})-(\ref{asum_2}) into (\ref{grav_eom}), the field equations for the metric become
\begin{align}
    R_{\mu\nu}=\dfrac{\xi}{2}\Big[\nabla_{\alpha}\nabla_{\mu}(b^{\alpha}b_{\nu})+\nabla_{\alpha}\nabla_{\nu}(b^{\alpha}b_{\mu})-\nabla^2(b_{\mu}b_{\nu})\Big].
\end{align}

The RHS of this equation can be rewritten in a simpler form
\begin{align}
    R_{\mu\nu}=\dfrac{\xi}{2}\nabla_{\alpha}\big[b^{\alpha}(\nabla_{\mu}b_{\nu}+\nabla_{\nu}b_{\mu})\big].\label{Ricci_eq}
\end{align}

This set of equations is much easier to solve than the initial one.

To proceed further, let us investigate the condition $b_{\mu\nu}=2\nabla_{[\mu}b_{\nu]}=0$. This condition is much easier to investigate in terms of $p$--forms. If one introduces the 1--form $\mathbf{b}=b_{\mu}\dd x^{\mu}$, then the condition $b_{\mu\nu}=0$ can be written as
\begin{align}
    \dd\mathbf{b}=0,
\end{align}
meaning that $\mathbf{b}$ is closed. Now one can use the Poincar\'{e} lemma that says that if on any open subset of a given manifold $\mathcal{M}$ the $p$--form $\omega$ is closed, then there exists a $(p-1)$--form $\rho$ such that $\omega=\dd\rho$. In our case, $\omega$ is the one-form $\mathbf{b}$ and $\rho$ is a scalar function. Thus at any open subset $\mathcal{U}$ of $\mathcal{M}$, the bumblebee field is given by
\begin{align}
    \mathbf{b}=b\, \dd\rho,~~~~~\rho\in\mathcal{FU}.
\end{align}
(Notice that to the RHS of this condition, we added a constant $b$ such that the normalization condition will be easily satisfied).

However, there exists a question of how to find $\rho$ and relate it to the metric functions. For this, we employ the normalization condition $b^{\mu}b_{\nu}=\epsilon b^2$. This condition means that $b_{\mu}$ is proportional to some spacelike (if $\epsilon=1$) or timelike (if $\epsilon=-1$) tetrad 1--form. Thus, there exists a tetrad covector, proportional to $\mathbf{b}$. The corresponding tetrad covector is given by
\begin{align}
    \mathbf{e}=\dd\rho.
\end{align}

This means that the metric and the bumblebee field are given by
\begin{align}
    ds^2=\epsilon\, \dd\rho^2+\gamma_{ij}\,dx^idx^j,~~~~~\mathbf{b}=b\,\dd\rho.\label{gaus_norm}
\end{align}

Thus, we have shown that employing the condition of normalization of the bumblebee field ($b^{\mu}b_{\mu}=\epsilon b^2$) and the condition of absence of the Faraday-like tensor ($b_{\mu\nu}=0$) requires the spacetime (even without employing the field equations) to have a possibility of being written in the Gaussian normal coordinates (\ref{gaus_norm}). Further analysis is thus significantly simplified, as in these coordinates it is easier to analyze the field equations.

This specific form of the metric suggests employing the standard expressions for the Ricci tensor for 3+1 splitting and analyzing equations (\ref{Ricci_eq}) using them. First of all, we notice that the Christoffel symbols for the metric (\ref{gaus_norm}) are given by (here and further the indices $i,~j,~k$ are any spacetime indices, except for $\rho$)
\begin{align}
    \Gamma^{\rho}_{\rho\rho}=0,~~\Gamma^{i}_{\rho\rho}=0,~~\Gamma^{\rho}_{\rho i}=0,~\Gamma^{\rho}_{ij}=-\epsilon K_{ij},~~\Gamma^{i}_{\rho j}=K^{i}_{j},~~\Gamma^{i}_{jk}=^{(3)}\Gamma^{i}_{jk},
\end{align}
where 
\begin{align}
    K_{ij}=\dfrac{1}{2}\partial_{\rho}g_{ij}
\end{align}
is an extrinsic curvature. The Ricci tensor components are given by (see \cite{Cook2000,Gourgoulhon2004} and our Supplementary Material)
\begin{align}
    R_{\rho\rho}=&\,-\partial_\rho K-K_{i j}K^{ij},\nonumber\\
    R_{\rho i}=&\,\nabla^{j}K_{ji}-\nabla_i K,\label{Ricci_tens_comp}\\
    R_{ij}=&\,^{(3)}R_{ij}-\epsilon(\partial_\rho K_{ij}-2 K_{ik}K^k_j+K K_{ij}).\nonumber
\end{align}

Now let us investigate various components of the RHS of (\ref{Ricci_eq}). First of all, we notice that $\nabla_\rho b_{\rho}=\partial_{\rho}b_{\rho}-\Gamma^{\rho}_{\rho\rho}b_{\rho}$. But as $b_{\rho}=b=\mathrm{const}.$ and as $\Gamma^{\rho}_{\rho\rho}=0$, we see that the $\rho\rho$ component of the RHS of (\ref{Ricci_eq}) is zero. Also, we see that $\nabla_{i}b_{\rho}=\partial_{i}b_{\rho}-\Gamma^{\rho}_{i\rho}b_{\rho}=0$ and $\nabla_{\rho}b_{i}=\partial_{\rho}b_{i}-\Gamma^{\rho}_{\rho i}b_{\rho}=0$, because $b_{\rho}=b=\mathrm{const}.$, $b_{i}=0$, and $\Gamma^{\rho}_{i \rho}=0$. This means that the $\rho i$ and $i \rho$ components of the RHS of (\ref{Ricci_eq}) are also zero. By the way, the fact that these two components are zero means that the bumblebee field equation (\ref{bb_eq1}) is automatically satisfied.

The next step is to calculate the $i j$ components. For this, we notice that 
\begin{align}
    \nabla_i b_{j}=\partial_{i}b_{j}-\Gamma^{\rho}_{ij}b_{\rho}=\epsilon b K_{i j}.\label{cov_d_b}
\end{align}

Thus the $i j$ components of the RHS of (\ref{Ricci_eq}) become
\begin{align}
    \dfrac{\xi}{2}\nabla_{\alpha}\big[b^{\alpha}(\nabla_{\mu}b_{\nu}+\nabla_{\nu}b_{\mu})\big]=\epsilon\xi b \,\nabla_{\alpha}(b^{\alpha} K_{ij})=\epsilon\xi b\big[ (\nabla_{\alpha}b^{\alpha})K_{i j}+ \,b^{\alpha}\nabla_{\alpha}(K_{ij})\big].
\end{align}

By taking trace of (\ref{cov_d_b}) with respect to $\gamma^{ij}$, we can see that $\nabla_{\alpha}b^{\alpha}=\epsilon b K$. Also, we use the fact that the covariant derivative of extrinsic curvature is
\begin{align}
    b^{\alpha}\nabla_{\alpha}(K_{ij})=\epsilon b\nabla_{\rho}(K_{ij})=\epsilon b (\partial_{\rho}K_{ij}- \Gamma^{k}_{\rho i}K_{k j}-\Gamma^{k}_{\rho j}K_{k i})=\epsilon b(\partial_{\rho}K_{ij}-2 K_{ik}K^{k}_{j}).
\end{align}

Substituting these expressions, we obtain that the $i j$ components of the RHS of (\ref{Ricci_eq}) are
\begin{align}
   \dfrac{\xi}{2}\nabla_{\alpha}\big[b^{\alpha}(\nabla_{\mu}b_{\nu}+\nabla_{\nu}b_{\mu})\big]= \xi b^2(\partial_{\rho}K_{ij}-2 K_{ik}K^{k}_{j}+ K K_{ij}).
\end{align}

Thus, putting this expression into equation (\ref{Ricci_eq}) and using (\ref{Ricci_tens_comp}), we see that the field equations take the form:
\begin{align}
    -\partial_\rho K-K_{i j}K^{ij}&=0,\nonumber\\
    \nabla^{j}K_{ji}-\nabla_i K&=0,\label{fin_set}\\
    ^{(3)}R_{ij}-(\epsilon+\xi b^2)(\partial_{\rho}K_{ij}-2 K_{ik}K^k_j+K K_{ij})&=0.\nonumber
\end{align}

In principle, one can solve this set of equations to find the metric $\gamma_{ij}(\rho, x^i)$. However, we notice that \textit{there exists a quite simple transformation that brings this set of equations into vacuum field equations}. Indeed, notice that if we in the metric functions $\gamma_{ij}$, perform a relabeling of the argument
\begin{align}
    \gamma_{ij}(\rho, x^i)\to \tilde{\gamma}_{ij}=\gamma_{i j}\Big(\dfrac{\rho}{\sqrt{1+\epsilon \xi b^2}},x^i\Big),
\end{align}
then $K_{ij}\to \dfrac{1}{\sqrt{1+\epsilon \xi b^2}} \tilde{K}_{ij}$, and the set (\ref{fin_set}) become
\begin{align}
    -(\partial_\rho \tilde{K}+\tilde{K}_{i j}\tilde{K}^{ij})&=0,\\
    \nabla^{j}\tilde{K}_{ji}-\nabla_i \tilde{K}&=0,\\
    ^{(3)}\tilde{R}_{ij}-\epsilon(\partial_\rho \tilde{K}_{ij}-2 \tilde{K}_{ik}\tilde{K}^k_j+\tilde{K} \tilde{K}_{ij})&=0.
\end{align}

This set of equations is the vacuum field equations for $\tilde{\gamma}_{ij}$ with $b=0$, and this means that by applying such a transformation on the metric satisfying the bumblebee field equations, one obtains a metric satisfying the vacuum field equations. This is quite an important result, namely, if one starts with the vacuum metric, written in the Gaussian normal coordinates
\begin{align}
    ds^2=\epsilon\, \dd\rho^2+\gamma_{ij}(\rho, x^i)\,\dd x^i\dd x^j,\label{Gaus_norm_coords}
\end{align}
then the metric 
\begin{align}
    ds_b^2=\epsilon \,\dd\rho^2+\gamma_{ij}\Big(\dfrac{\rho}{\sqrt{1+\epsilon \xi b^2}},x^i\Big)\,\dd x^i \dd x^j\label{resc_metr}
\end{align}
will satisfy the bumblebee field equations with the bumblebee field $\mathbf{b}=b d\rho$. 

Practically, as it will be clear, it is better to work with a slightly modified form of the metric that is achieved by the transformation $\rho\to \sqrt{1+\epsilon \xi b^2}\rho$, so that the metric takes the form:
\begin{align}
    ds_b^2=\epsilon(1+\epsilon \xi b^2)\,\dd\rho^2+\gamma_{ij}(\rho,x^i)\,\dd x^i\dd x^j,\label{Final_transform_metr}
\end{align}
with the bumblebee field 
\begin{align}
    \mathbf{b}=b\sqrt{1+\epsilon\xi b^2}\,\dd\rho.\label{Final_transform_bb}
\end{align}

Thus, to summarize, the generating technique can be formulated as follows:
\begin{itemize}
    \item Take any spacetime satisfying the vacuum Einstein equations.
    \item Perform some transformation $T$ of coordinates that brings the metric to the form written in the Gaussian normal coordinates (\ref{Gaus_norm_coords}).
    \item Change the metric to (\ref{Final_transform_metr}). This spacetime will satisfy both the metric and the bumblebee field equations, with the bumblebee field given by (\ref{Final_transform_bb}).
    \item For a better presentation of the resulting metric (\ref{Final_transform_metr}), one may perform an inverse transformation $T^{-1}$ of coordinates.
\end{itemize}

\subsection{Relation to the Hamilton--Jacobi equation}

The generating technique, presented in the previous subsection, does not seem to be complicated if one is able to transform the metric to the form of Gaussian normal coordinates. However, even though this seems to be an easy task, technically, it is not trivial. In this subsection, we will show how one can achieve such coordinates.

For this, let us assume that we have a spacetime such that in the coordinates $y^{\mu}$ it takes the form $ds^2=\tilde{g}_{\mu\nu}dy^{\mu}dy^{\nu}$. Then we perform a transformation of coordinates to $(\rho, x^i)$. During this transformation, the $g^{\rho\rho}$ component of the metric becomes
\begin{align}
    g^{\rho\rho}=\tilde{g}^{\mu\nu}\dfrac{\partial\rho}{\partial y^{\mu}}\dfrac{\partial\rho}{\partial y^{\nu}}.
\end{align}

From the other side, as we wish to obtain the Gaussian normal coordinates (\ref{Gaus_norm_coords}), this metric component has to be equal to $\epsilon$. This means that the normal coordinate $\rho(y^\mu)$ has to satisfy the Hamilton--Jacobi equation 
\begin{align}
    \tilde{g}^{\mu\nu}\dfrac{\partial \rho}{\partial y^\mu}\dfrac{\partial \rho}{\partial y^\nu}=\epsilon.\label{HJ_eq}
\end{align}

If this is done, then the bumblebee field and the metric, satisfying the bumblebee gravity field equations, are given by
\begin{align}
    ds_b^2=\Big(\tilde{g}_{\mu\nu}+\dfrac{\xi}{1+\epsilon \xi b^2}b_{\mu}b_{\nu} \Big)\,\dd y^{\mu}\dd y^{\nu},~~~~~    \mathbf{b}=b\sqrt{1+\epsilon \xi b^2}\,\dd \rho\label{Final_metr_and_bb}
\end{align}
This expression for the metric can be checked by direct substitution of the 1-form $\mathbf{b}$, and comparing the resulting expression with (\ref{Final_transform_metr}).

This proposes the \textit{final generating algorithm}:
\begin{itemize}
    \item Take a background spacetime $\tilde{g}_{\mu\nu}$, satisfying the vacuum Einstein field equations.
    \item Solve the Hamilton--Jacobi equation (\ref{HJ_eq}) on this spacetime for the function $\rho$.
    \item Using $\rho$, generate the metric and the bumblebee field using the expressions (\ref{Final_metr_and_bb}). They will satisfy all the bumblebee gravity field equations.  
\end{itemize}

Thus, if one wants to summarize this generating technique, its most important feature is that one can take a vacuum spacetime, consider some geodesically moving observer in it, and associate a bumblebee field with this observer that is codirected with the tangent vector to this geodesic. Then one needs to modify the metric, adding a term with the tensor product of this tangent vector. 

Notice that this result is a step forward compared to the condition (18) in \cite{Poulis2022}. Equation (18) from \cite{Poulis2022} means that the bumblebee field is geodesic \textit{on the resulting spacetime} $g_{\mu\nu}$, however, here we show that the bumblebee field can be found as the geodesic vector field on the \textit{seed} spacetime $\tilde{g}_{\mu\nu}$, thus significantly reducing the complexity of finding the bumblebee field.

This generating technique thus explicitly uses the defining property of the bumblebee gravity, namely the Lorentz violation. We modify the metric by adding new terms related to a specific direction in the spacetime, thus directly introducing Lorentz-violating terms to the metric. Also, we wish to remark that this generating technique is somehow analogous to the one developed in the metric-affine bumblebee gravity \cite{Filho2023,ArajoFilho2024}, but the explicit coefficients are different.

Even though on the level of an idea this generating technique is quite simple, its application to some known spacetimes is still complicated as it requires knowing the solution to the Hamilton--Jacobi field equations (or at least the derivatives of $\rho$ with respect to \textit{all} the coordinates as, in fact, only the derivatives are required in (\ref{Final_metr_and_bb})). We postpone the application of this method for some specific spacetimes to Sec. \ref{Sec_applic}, and now we move to the analysis of its application to other theories.

\section{Generalization to other theories}\label{Sec_generalizations}

In this section, we wish to show that this generating technique does not work solely for the case of vacuum, but also can be generalized to the case of non-zero cosmological constant, and a presence of the electromagnetic field. We start with the case when only the cosmological constant is present.

\subsection{Non-zero cosmological constant}

In this case, the equations governing the gravitational and bumblebee fields are \eqref{grav_eom} and \eqref{bb_eom}, however, we do not need to assume $\Lambda=0$. Moreover, we would like to note that previously, while considering the case with zero cosmological constant, we assumed that for our solution $V=0$ and $V'=0$. In the cases with the non-zero cosmological constant, as was shown in \cite{Maluf2021}, one cannot use this assumption, and the effective potential has to be assumed in the form $V=\dfrac{\lambda}{2}X$. Moreover, to have consistent field equations, the constant $\lambda$ has to be chosen to be
\begin{align}
    \lambda=\Lambda \dfrac{\xi}{\kappa}\dfrac{1}{1+\epsilon \xi b^2}.\label{lambda_spec}
\end{align}

Substituting this expression for the effective potential into the field equation (\ref{bb_eom}), one obtains (here as previously we assume $b_{\mu\nu}=0$)
\begin{align}
    b_{\mu}R^{\mu\nu}=\dfrac{\kappa \lambda}{\xi}\,b^{\nu}.\label{br_lambda}
\end{align}

Using this relation and substituting it into the field equations (\ref{grav_eom}), one obtains (employing (\ref{lambda_spec}))
\begin{align}
    R_{\mu\nu}-\dfrac{\xi}{2}\nabla_{\alpha}\big[b^{\alpha}(\nabla_{\mu}b_{\nu}+\nabla_{\nu}b_{\mu})\big]=\Lambda\Big(g_{\mu\nu}-\dfrac{\xi}{1+\epsilon \xi b^2}b_{\mu}b_{\nu}\Big).\label{lambda_bb_eom}
\end{align}

Now let us employ the generating technique, described after (\ref{Final_metr_and_bb}), and inspect it. If we perform such a transformation, then the LHS of (\ref{lambda_bb_eom}), as was shown in the previous section, becomes $\tilde{R}_{\mu\nu}$, namely the Ricci tensor of the background metric. The RHS of (\ref{lambda_bb_eom}) after this transformation becomes $\Lambda \tilde{g}_{\mu\nu}$, and thus the field equation is given by
\begin{align}
    \tilde{R}_{\mu\nu}=\Lambda \tilde{g}_{\mu\nu}.
\end{align}

This equation is automatically satisfied because the ``seed" metric $\tilde{g}_{\mu\nu}$ satisfies the Einstein field equation with the cosmological constant. 

Uniqueness of the generating technique can be proven in a way analogous to the one considered in the previous section. Namely, for this, one has to write the Gaussian normal coordinates, perform the rescaling (\ref{resc_metr}), and see that both the LHS and the RHS of the  Einstein equations transform into vacuum Einstein equations with $\Lambda$.

\subsection{Non-zero electromagnetic field}

The next theory modification we wish to explore is the one that adds an electromagnetic field. Let us consider a generic, quadratic in the Faraday tensor $F_{\mu\nu}$, action coupled with the bumblebee field $B_{\mu}$:
\begin{align}
    S_m=\int \sqrt{-g}\,d^4 x\Big[\Big(-\dfrac{1}{4}+\dfrac{\gamma_1}{2\kappa} B_{\lambda}B^{\lambda}\Big) F_{\mu\nu}F^{\mu\nu}+\dfrac{(\gamma_2+\gamma_3 B_{\lambda}B^{\lambda})}{2\kappa} B_{\mu}F^{\mu\nu}B^{\rho}F_{\rho\nu}\Big].\label{S_matter}
\end{align}

Such a theory with $\gamma_2=0=\gamma_3$ was considered in \cite{Liu2025}, while the theory with $\gamma_3=0$ was considered in \cite{Li2025_2}. It was shown that both these theories allow for spherically symmetric solutions (when the rest of the non-zero $\gamma$'s are fixed to some specific values), moreover, the theory considered in \cite{Li2025_2} allows for dyonic black holes. In this section, we consider a more general action, and as will be clear from the analysis below, such a generalized theory is worth considering because it allows to apply the generating technique discussed in previous sections without the need to solve the field equations\footnote{Notice that the additional term with $\gamma_3$ is of the 4-th order in $B^\mu$. Because of this, one does not expect it to give any considerable modification because the combination $\xi b^2$ is considered to be small. Anyway, for the exact generating technique, this term is required even though at the observational scales it does not play any role.}.

By varying the total action (\ref{bb_act}) with the matter action (\ref{S_matter}) with respect to the metric, one obtains such field equations
\begin{align}
    &R_{\mu\nu}-\Lambda g_{\mu\nu}=\kappa\Big[V'(2 B_{\mu}B_{\nu}-B_{\rho}B^{\rho} g_{\mu\nu})+B_{\mu}^{~\alpha}B_{\nu\alpha}+V g_{\mu\nu}-\dfrac{1}{4}B_{\alpha\beta}B^{\alpha\beta}g_{\mu\nu}\Big]+\dfrac{\xi}{4}g_{\mu\nu}\nabla^2(B^{\alpha}B_{\alpha})\nonumber\\
    &+\xi\Big[\dfrac{1}{2}B^{\alpha}B^{\beta}R_{\alpha\beta}g_{\mu\nu}-2B_{(\mu}R_{ \nu)\alpha}B^{\alpha}\Big]+\dfrac{\xi}{2}\Big[\nabla_{\alpha}\nabla_{\mu}(B^{\alpha}B_{\nu})+\nabla_{\alpha}\nabla_{\nu}(B^{\alpha}B_{\mu})-\nabla^2(B_{\mu}B_{\nu})\Big]\nonumber\\
    &+(\kappa-2\gamma_1 B_{\lambda}B^{\lambda})\Big[F_{\mu}^{~\rho}F_{\nu\rho}-\dfrac{1}{4}g_{\mu\nu}F_{\rho\sigma}F^{\rho\sigma}\Big]+\dfrac{\gamma_1}{2} (B_{\rho}B^{\rho} g_{\mu\nu}-2B_{\mu}B_{\nu})F_{\rho\sigma}F^{\rho\sigma}\nonumber\\
    &+\dfrac{\gamma_3}{2}B_{\lambda}F^{\lambda\rho}B^{\sigma}F_{\sigma\rho} (B_{\delta}B^{\delta} g_{\mu\nu}-2B_{\mu}B_{\nu})\label{grav_eom_2}\\&+(\gamma_2+B_{\delta}B^{\delta} \gamma_3)(B_{\lambda}F^{\lambda\rho}B^{\sigma}F_{\sigma\rho} g_{\mu\nu}-2B^{\rho}B_{(\mu}F_{\nu)}^{~\sigma}F_{\rho\sigma}-B^{\rho}B^{\sigma}F_{\mu\rho}F_{\nu\sigma}),\nonumber
\end{align}
where $(\mu\nu)$ is the symmetrization with respect to the corresponding indices.

Varying the action with respect to the bumblebee field, one obtains:
\begin{align}
    &\dfrac{1}{\kappa}\Big(\xi B_{\mu}R^{\mu\nu} +B^{\nu}(\gamma_1 F_{\rho\sigma}F^{\rho\sigma}+\gamma_3 B^{\rho}F_{\rho\lambda}B^{\sigma}F_{\sigma}^{~\lambda})+(\gamma_2+B_{\lambda}B^{\lambda} \gamma_3)B^{\rho}F_{\rho\sigma}F^{\nu\sigma}\Big)\nonumber\\&+\nabla_{\mu}B^{\mu\nu}-2 V'B^{\nu}=0.\label{bb_eom_2}
\end{align}

The last equation can be obtained by the variation of the 4-potential $A_{\mu}$ that gives:
\begin{align}
    \nabla_{\mu}\Big[(\kappa-2 \gamma_1 B_{\lambda}B^{\lambda})F^{\mu\nu}+2(\gamma_2+ \gamma_3 B_{\lambda}B^{\lambda} )B_{\rho}F^{\rho[\mu}B^{\nu]}\Big]=0,
\end{align}
where $[\mu\nu]$ is the antisymmetrization with respect to the corresponding indices.

The resulting theory seems to be extremely complicated. However, we can significantly simplify the corresponding metric field equations (\ref{grav_eom_2}) if we employ the same assumptions as in the previous sections. Namely, if we assume that the bumblebee field $B^{\mu}$ is given by its VEV $b_{\mu}$, if we assume that the ``Faraday" tensor for the bumblebee field vanishes ($b_{\mu\nu}=0$), and if we assume the potential $V=\dfrac{\lambda}{2}X$, where $\lambda$ is given by (\ref{lambda_spec}). Then (\ref{grav_eom_2}) becomes (after employing (\ref{bb_eom_2})):
\begin{align}
    &R_{\mu\nu}=\Lambda\Big(g_{\mu\nu}-\dfrac{\xi}{1+\epsilon \xi b^2}b_{\mu}b_{\nu}\Big)+\dfrac{\xi}{2}\nabla_{\alpha}[b^{\alpha}(\nabla_{\mu}b_{\nu}+\nabla_{\nu}b_{\mu})]\nonumber\\
    &+(\kappa-2\epsilon \gamma_1 b^2)\Big[F_{\mu}^{~\rho}F_{\nu\rho}-\dfrac{1}{4}g_{\mu\nu}F_{\rho\sigma}F^{\rho\sigma}\Big]+\gamma_1 B_{\mu}B_{\nu}F_{\rho\sigma}F^{\rho\sigma}+\gamma_3 (B_{\lambda}F^{\lambda\rho}B^{\sigma}F_{\sigma\rho}) b_{\mu}b_{\nu}\label{grav_eom_3}\\
    &+\dfrac{(\gamma_2+\epsilon b^2 \gamma_3)}{2}(B_{\lambda}F^{\lambda\rho}B^{\sigma}F_{\sigma\rho} g_{\mu\nu}-2B^{\rho}B^{\sigma}F_{\mu\rho}F_{\nu\sigma}),\nonumber
\end{align}
with the Maxwell equation
\begin{align}
    \nabla_{\mu}\Big[(\kappa-2\epsilon \gamma_1 b^2)F^{\mu\nu}+2(\gamma_2+\epsilon \gamma_3 b^2 )B_{\rho}F^{\rho[\mu}B^{\nu]}\Big]=0.
\end{align}

In general, this theory is still quite complicated and it is not guaranteed that for a generic $\gamma_1,~\gamma_2$ and $\gamma_3$ there may exist exact solutions. However, we wish to show that for a specific choice of $\gamma$s the corresponding theory significantly simplifies, and the generating technique, developed in previous sections, can be applied even in this case.

The specific choice of coefficients mentioned above is: 
\begin{align}
    \gamma_1=\dfrac{\kappa \xi}{2(2+3\epsilon \xi b^2)},~~\gamma_2=-\kappa \xi,~~\gamma_3=2\xi \gamma_1.\label{coefs}
\end{align}

If one chooses such coefficients, the corresponding field equations become:
\begin{align}
    R_{\mu\nu}=&\Lambda\Big(g_{\mu\nu}-\dfrac{\xi}{1+\epsilon \xi b^2}b_{\mu}b_{\nu}\Big)+\dfrac{\xi}{2}\nabla_{\alpha}[b^{\alpha}(\nabla_{\mu}b_{\nu}+\nabla_{\nu}b_{\mu})]\nonumber\\
    +&\dfrac{\kappa}{2(2+3\epsilon \xi b^2)}\Big[(1+\epsilon\xi b^2)(4 F_{\mu\rho}F_{\nu}^{~\rho}-g_{\mu\nu}F_{\rho\sigma}F^{\rho\sigma})\\
    +&2 \xi(1+\epsilon \xi b^2) b^{\rho}b^{\sigma}(2 F_{\rho\mu}F_{\sigma \nu}-F_{\rho}^{~\lambda}F_{\sigma\lambda}g_{\mu\nu})+\xi b_{\mu}b_{\nu}(F_{\rho\sigma}F^{\rho\sigma}+2\xi b^{\rho}b^{\sigma}F_{\rho\lambda}F_{\sigma}^{~\lambda})\Big],\nonumber
\end{align}
with the Maxwell equation:
\begin{align}
    \nabla_{\mu}(F^{\mu\nu}-2\xi b_{\rho}F^{\rho[\mu}b^{\nu]})=0.
\end{align}

The basic observation about this theory is that if we define a new tensor $\tilde{F}^{\mu\nu}$ as
\begin{align}
    \tilde{F}^{\mu\nu}=\sqrt{2}\sqrt{\dfrac{1+\epsilon \xi b^2}{2+3\epsilon \xi b^2}}(F^{\mu\nu}-2\xi b_{\rho}F^{\rho[\mu}b^{\nu]}),\label{em_rel}
\end{align}
then the Maxwell equations become $\nabla_{\mu}\tilde{F}^{\mu\nu}=0$.

Moreover, the field equations become simply
\begin{align}
    R_{\mu\nu}-&\dfrac{\xi}{2}\nabla_{\alpha}[b^{\alpha}(\nabla_{\mu}b_{\nu}+\nabla_{\nu}b_{\mu})]=\Lambda h_{\mu\nu}\nonumber\\
    +&\kappa\Big[h_{\mu \rho}h_{\sigma\lambda}h_{\nu\delta}\tilde{F}^{\mu \sigma}\tilde{F}^{\delta\lambda}-\dfrac{1}{4} h_{\mu\nu}(\tilde{F}^{\rho\sigma}\tilde{F}^{\delta\lambda}h_{\rho\delta}h_{\sigma\lambda})\Big],\label{fin_eq_em}
\end{align}
where we for shortness defined $h_{\mu\nu}\equiv g_{\mu\nu}-\dfrac{\xi}{1+\epsilon \xi b^2}b_{\mu}b_{\nu}.$

Now let us employ the generating technique (\ref{Final_metr_and_bb}), described in previous sections. As was shown in Sec. \ref{Sec_der}, the LHS of this equation becomes the Ricci tensor of the background metric. The term with the cosmological constant becomes $\Lambda \tilde{g}_{\mu\nu}$, while the novel additional term is precisely the usual energy-momentum tensor of the electromagnetic field. Thus, for this specific theory, we were able to show that the field equations take the form of 
\begin{align}
    \tilde{R}_{\mu\nu}=\Lambda \tilde{g}_{\mu\nu}+\kappa\Big[\tilde{F}_{\mu\rho}\tilde{F}_{\nu}^{\rho}-\dfrac{1}{4} \tilde{g}_{\mu\nu}\tilde{F}_{\rho\sigma}\tilde{F}^{\rho\sigma}\Big],
\end{align}
here lowering and raising of indices are considered to be with respect to the background metric $\tilde{g}_{\mu\nu}$. The Faraday tensor $\tilde{F}^{\mu\nu}$ satisfies the usual Maxwell equations:
\begin{align}
    \nabla_{\mu}\tilde{F}^{\mu\nu}=0,
\end{align}
which is exactly the Einstein--Maxwell theory with $\Lambda$\footnote{Also, we wish to notice that if one contracts the (\ref{fin_eq_em}) with $b^{\mu}$, then one obtains the equation for the bumblebee field (\ref{bb_eom_2}) with the coefficients, given by (\ref{coefs}). Thus, this generating technique also automatically satisfies the bumblebee field equation. For the proof, see Supplementary material.}.

One may ask a question about how to obtain the Faraday tensor and the 4-potential such that the condition (\ref{em_rel}) is satisfied. For this we first of all notice that, if one contracts (\ref{em_rel}) with $b_{\mu}$, then one obtains a relation
\begin{align}
    b_{\mu}\tilde{F}^{\mu\nu}=\sqrt{2}\sqrt{\dfrac{1+\epsilon \xi b^2}{2+3\epsilon \xi b^2}} (1+\epsilon \xi b^2)b_{\rho}F^{\rho\nu}.
\end{align}

Using this relation and (\ref{em_rel}), one can determine $F^{\mu\nu}$ in terms of $\tilde{F}^{\mu\nu}$:
\begin{align}
    F^{\mu\nu}=\sqrt{\dfrac{2+3\epsilon \xi b^2}{2+2\epsilon\xi b^2}}\Big(\tilde{F}^{\mu\nu}+\dfrac{2\xi}{1+\epsilon \xi b^2}b_{\rho}\tilde{F}^{\rho[\mu}b^{\nu]}\Big).\label{Faraday_rel}
\end{align}

Also, by lowering indices, one obtains
\begin{align}
    F_{\mu\nu}=\sqrt{\dfrac{2+3\epsilon \xi b^2}{2+2\epsilon\xi b^2}}\tilde{g}_{\mu\rho}\tilde{g}_{\nu\sigma}\tilde{F}^{\rho\sigma}
\end{align}

Here on the LHS, the indices are lowered with respect to the metric $g_{\mu\nu}$, while on the RHS with respect to $\tilde{g}_{\mu\nu}$. At the end, we see that the novel 2-form $\mathbf{F}$ and the old 2-form $\tilde{\mathbf{F}}$ are related by simple rescaling. This means that the 4-potential $\mathbf{A}$, generating the 2-form $\mathbf{F}$, is simply given by
\begin{align}
    \mathbf{A}=\sqrt{\dfrac{2+3\epsilon \xi b^2}{2+2\epsilon\xi b^2}}\tilde{\mathbf{A}}\label{A_rel}
\end{align}
Because of this, we notice that as $\dd \tilde{\mathbf{F}}=0$, then $\dd \mathbf{F}=0$ (as the external derivative of a form is independent of the metric), what means that $\mathbf{F}$ is closed as well as $\tilde{\mathbf{F}}$.

This allows us to formulate the generating algorithm in the presence of the cosmological constant and the electromagnetic field in such a way:
\begin{itemize}
    \item Take a background metric $\tilde{g}_{\mu\nu}$ and the corresponding Faraday tensor $\tilde{F}^{\mu\nu}$, satisfying the Einstein-Maxwell field equations (possibly with $\Lambda$).
    \item Solve the Hamilton--Jacobi equation (\ref{HJ_eq}) on this spacetime for the function $\rho$.
    \item Using $\rho$, generate the metric $g_{\mu\nu}$, the bumblebee field $b_{\mu}$, the 4-potential $A_{\mu}$ and the Faraday tensor $F^{\mu\nu}$ using the expressions (\ref{Final_metr_and_bb}), (\ref{Faraday_rel}) and (\ref{A_rel}). They will satisfy all the bumblebee gravity field equations for the theory given by the action (\ref{bb_act}) and (\ref{S_matter}), with the coefficients (\ref{coefs}).
\end{itemize}

The uniqueness of this generating technique is straightforward and can be proven in a way analogous to the one presented in the previous sections by writing the metric in the Gaussian normal coordinates. We will thus not repeat this calculation here.

The advantage of our new theory, considered above, in comparison with theories investigated in \cite{Liu2025} and \cite{Li2025_2} is that, by employing the developed in \cite{Poulis2022} generating technique, this theory \textit{possesses an analog of any electrovaccum spacetime}. This allows bypassing the issue that, as in the theory considered in \cite{Liu2025}, only electrically charged spacetimes exist, while in the theory considered in \cite{Li2025_2} this issue was bypassed; however, only a spherically symmetric solution and a solution analogous to the charged Taub-NUT spacetime were found within these models. This generating technique guarantees that \textit{any} electrovacuum spacetime has its ``bumblebee analog" in our theory.

We also wish to comment on other theories to which this technique, in principle, can be applied. Notice that for all the theories we have considered so far, employing the generating technique brought the field equations to the form of the field equations without the bumblebee field. This, in particular, means that if one has a more general theory incorporating other physical fields, then this generating technique will work if, under this technique, the energy-momentum tensor of such a theory transforms to the energy-momentum tensor without the bumblebee field. Understanding this interesting property in other theories is postponed for future work.

Now, when we have proven the uniqueness of this generating technique and shown that it also works in the case with the cosmological constant and the electromagnetic field, let us apply it to various electrovacuum solutions known in General Relativity. 

\section{Application of the generating technique}\label{Sec_applic}

In this section, we apply the generating technique investigated so far, and see what spacetimes we are able to achieve. Also, instead of just presenting the solutions, we elaborate a lot on finding the conditions under which the bumblebee field is \textit{globally real} and on its physical interpretations. 

\subsection{The most general spacetime that allows for the separation: Kerr--Newman--Taub-NUT--(A)dS}\label{Sec_gen_metr}

As the generating technique discussed above requires a solution to the Hamilton--Jacobi (HJ) equation, the basic idea is to try to apply it to the cases when the HJ equation has an analytical solution (at least in the form of indefinite integrals). These are the cases when the HJ equation is separable. As was shown in \cite{Krtou2008}, the most general vacuum spacetime, possessing a Killing tensor (responsible for the separation of the HJ equation), is the Kerr--Taub-NUT--(A)dS spacetime. If we in addition allow for an electromagnetic field, then the most general solution is the Kerr-Newman-Taub-NUT--(A)dS spacetime (see Sec. 3.2.2 in \cite{Frolov2017}). Review of the physical properties and various coordinate representations of these spacetimes can be found in \cite{Podolsk2021,Podolsk2023,Ovcharenko2025_2,Ovcharenko2024}. 

The corresponding metric of Kerr--Newman--Taub-NUT--(A)dS is given by (we use the parametrization presented in \cite{Podolsk2023})
\begin{align}
    ds^2=&-\dfrac{Q}{\rho^2}\big(\dd t-(a(1-x^2)+2l (1-x))\dd\varphi\big)^2+\dfrac{\rho^2}{Q}\dd r^2\nonumber\\
    &+\dfrac{\rho^2}{P}\dd x^2+\dfrac{P}{\rho^2}\big(a \dd t-(r^2+(a+l)^2)\dd\varphi\big)^2.\label{ds2_Kerr-NUT}
\end{align}
where $\rho^2=r^2+(a x+l)^2$, and the two metric functions $Q$ and $P$ are given by
\begin{align}
    Q&=a^2+q^2-l^2-2m r+r^2-\dfrac{\Lambda}{3}r^2\Big(r^2+a^2+3 l^2\Big),\label{Q_gen}\\
    P&=(1-x^2)\Big(1+2\dfrac{\Lambda }{3}l(l+a x)+\dfrac{\Lambda}{3}(l+a x)^2\Big),\label{P_gen}
\end{align}
where $q$ is the charge of the BH, $a$ and $l$ are the Kerr and Taub-NUT twist parameters, $m$ is the mass parameter, $\Lambda$ is the cosmological constant. Notice that, instead of the polar coordinate $\theta$, we introduced a coordinate $x=\cos\theta$, because it is easier to work with this coordinate, as will be clear from the analysis below.

Now let us solve the Hamilton-Jacobi equation (\ref{HJ_eq}) in this spacetime. As the metric has two Killing vectors $\partial_t$ and $\partial_{\varphi}$, the function $\rho$ has such a form:
\begin{align}
    \rho=-E t+s(r,x)+L\varphi.\label{rho_gen}
\end{align}

Substituting this ansatz into (\ref{HJ_eq}), one obtains
\begin{align}
    &Q (\partial_r s)^2-\dfrac{(E((a+l)^2+r^2)-aL)^2}{Q}-\epsilon r^2=\nonumber\\
    &\epsilon (a x+l)^2-P (\partial_x s)^2-\dfrac{\big(L-E\big[a (1-x^2)+2l (1-x)\big]\big)^2}{P}.
\end{align}

This equation is separable, $s=f(r)+h(x)$. Then both LHS and RHS of this equation have to be constant, and one can write:
\begin{align}
    f'(r)&=\sigma_r\sqrt{\dfrac{\epsilon r^2-C}{Q}+\dfrac{(a L-E(r^2+(a+l)^2))^2}{Q^2}},\label{Kerr-NUT_f}\\
    h'(x)&=\sigma_x\sqrt{\dfrac{\epsilon (ax+l)^2+C}{P}-\dfrac{(L-E(a(1-x^2)+2l(1-x)))^2}{P^2}},\label{Kerr-NUT_h}
\end{align}
where $C$ is the separation constant, while $\sigma_r,~\sigma_x=\pm1$.

These equations can be integrated to find $f$ and $h$ explicitly. However, we do not require the explicit form of $f$ and $h$, but rather only their derivatives. This can be easily seen from (\ref{Final_metr_and_bb}) because for this generating technique, we do not require $\rho$ explicitly, but rather only $d\rho$. Thus, using these expressions for $f$ and $g$, we can write that 
\begin{align}
    \mathbf{b}=&b\sqrt{1+\epsilon \xi b^2}\Big(-E\dd t+\sigma_r\sqrt{\dfrac{\epsilon r^2-C}{Q}+\dfrac{(a L-E(r^2+(a+l)^2))^2}{Q^2}}\dd r\nonumber\\
    &+\sigma_x\sqrt{\dfrac{\epsilon (ax+l)^2+C}{P}-\dfrac{(L-E(a(1-x^2)+2l(1-x)))^2}{P^2}}\dd x+L \dd\varphi\Big),\label{bb_genn}\\
    g_{\mu\nu}=&\tilde{g}_{\mu\nu}+\dfrac{\xi}{1+\epsilon \xi b^2}\,b_{\mu}b_{\nu},\label{metr_genn}
\end{align}
where $\sigma_r,\sigma_x=\pm1$, and $\tilde{g}_{\mu\nu}$ is the background Kerr--Newman--Taub-NUT--(A)dS

Thus, we see that this simple generating technique allowed for the easy generation of new solutions to the bumblebee gravity. In addition, we would like to notice that this novel metric does not depend solely on the combination $\xi b^2$, representing the strength of the bumblebee field, but also adds 3 new parameters, namely $E,~L,~C$. These parameters play the role of energy, angular momentum, and the Carter constant associated with the geodesic particle, generating the bumblebee field (furtheron we will call these parameters as energy, angular momentum, and a Carter constant, meaning exactly this interpretation). By varying these constants, one may obtain bumblebee fields, pointing in various directions and thus giving rise to different spacetimes. This is important to emphasize (as this point was not so detailed analyzed in the literature) that the introduction of the bumblebee field is not unique, and for the same ``seed" space-time, one can obtain several bumblebee generalizations.

To understand the role of these parameters better, let us consider several special cases. 

\subsection{Vacuum spacetimes (no charge $q=0$ and $\Lambda=0$)}

First of all, we limit ourselves to the case of vacuum spacetimes to understand the influence of the parameters $E,~L$, and $C$ on this simpler case, and then to add a non-trivial electromagnetic field and a cosmological constant. 

\subsubsection{Static case (no twist: $a=0,~l=0$)}\label{Sec_static_vac}

Let us start with the static case and the case when $E=L=C=0$ (as it gives the simplest spacetime). In this case
\begin{align}
    \mathbf{b}=\pm b\sqrt{1+\epsilon \xi b^2}\sqrt{\dfrac{\epsilon }{f}}\,\dd r,\label{Schw_bb}
\end{align}
where $f=Q/r^2=1-2m/r$.

The background metric is given by
\begin{align}
    ds^2_{\mathrm{Sch}}=-f \dd t^2+\dfrac{\dd r^2}{f}+r^2(\dd\theta^2+\sin^2\dd \varphi^2).
\end{align}
Substituting $\mathbf{b}$ to (\ref{Final_metr_and_bb}), one sees that the final metric takes the form:
\begin{align}
    ds^2_b=-f \dd t^2+\dfrac{(1+\epsilon \xi b^2)}{f}\dd r^2+r^2(\dd\theta^2+\sin^2\dd \varphi^2).\label{casana_metr}
\end{align}

This is exactly the metric, found in \cite{Casana2018}. However, we have to discuss one important thing here. Notice that even though the metric functions are real for any value of $r$, the bumblebee field (\ref{Schw_bb}) does not have this property. For example, if the bumblebee field is spacelike ($\epsilon=+1$), then the bumblebee field is real \textit{only} above the horizon ($r>2m$). If the bumblebee field is timelike ($\epsilon=-1$), then it is real only below the horizon ($0<r<2m$), while, simultaneously, purely radial bumblebee field cannot be real both above and below the horizon. There is nothing unexpected here because the radial coordinate becomes timelike below the horizon, and thus the bumblebee field also has to change from spacelike to timelike to remain real below the horizon. However, this rapid change in the causal structure of the bumblebee field is highly unexpected. Also, notice that this property is independent of the coordinates we used because it is related to the causal structure of the covector $b_{\mu}$. Nevertheless, we have set $E=L=C=0$ to obtain this solution, and one may hope that by considering cases with some of these parameters being non-zero, one can avoid this issue. 

For this, let us consider the case when $E=L=0$, $C\neq 0$. In this case, the bumblebee field (\ref{bb_genn}) becomes
\begin{align}
    \mathbf{b}=b\sqrt{1+\epsilon \xi b^2}\Bigg[\sigma_r\sqrt{\dfrac{\epsilon}{f}\Big(1-\dfrac{\epsilon C}{r^2}\Big)}\,\dd r+\sigma_x\sqrt{C}\,\dd\theta\Bigg].
\end{align}

The metric, corresponding to this bumblebee field, is given by
\begin{align}
    ds_b^2=&-f\dd t^2+\Big(1+\epsilon \xi b^2\big(1-\epsilon C/r^2\big)\Big)\dfrac{\dd r^2}{f}+2 \sigma_r \sigma_x\xi b^2\sqrt{\dfrac{\epsilon C}{f}\Big(1-\dfrac{\epsilon C}{r^2}\Big)}\dd r \dd \theta\nonumber\\
    &+r^2\Big((1+\xi b^2 C/r^2)\dd\theta^2+\sin^2\theta \dd\varphi^2\Big).\label{static_with_C}
\end{align}

Even though this spacetime was already found \cite{Poulis2022}, it is still worth spending some time here discussing its properties, because in \cite{Poulis2022} authors did not investigate the physical properties of this spacetime in great detail. First of all, notice that this spacetime is not diagonal anymore. This non-diagonal term is caused by the non-zero coefficient $C$. Thus, we can notice that if $\mathbf{b}$ contains several terms, then the metric will acquire non-diagonal terms. In principle, one can think that this non-diagonal term can vanish if one performs some transformation of the coordinates, or even bring the metric to the form (\ref{casana_metr}) (as was obtained in the case considered in detail in \cite{Poulis2022}, where authors, in our notation, considered a non-zero $E$, while $C=0=L$). We can surely say that the second conjecture is incorrect: this follows from the fact that the \textit{algebraic type} (that is independent of the coordinates used) \textit{of the metric (\ref{static_with_C}) is I}.

However, there exists an additional issue. This non-diagonal term may become imaginary, which is not expected to happen. This problem can be avoided if one chooses $C=4m^2/\epsilon$ (in fact, this is the only choice, giving a globally real metric, because then the roots of numerator and denominator coincide). Then the metric takes the form
\begin{align}
    ds_b^2=&-f\dd t^2+\Big(1+\epsilon \xi b^2\big(1-4m^2/r^2\big)\Big)\dfrac{\dd r^2}{f}+4\sigma_r \sigma_x m \xi b^2\sqrt{1+\dfrac{2 m}{r}}\dd r \dd\theta\nonumber\\
    &+r^2\Big((1+4\epsilon\xi b^2m^2 /r^2)\dd\theta^2+\sin^2\theta \dd\varphi^2\Big),
\end{align}
while the bumblebee field is given by
\begin{align}
    \mathbf{b}=b\sqrt{1+\epsilon \xi b^2}\Bigg[\sigma_r\sqrt{\epsilon\Big(1+\dfrac{2 m}{r}\Big)}\dd r+2\sigma_x m \dd\theta\Bigg].
\end{align}

The bumblebee field will be real only if $\epsilon=+1$, meaning that it has to be \textit{spacelike}. This choice is quite natural as it involves only radial and polar coordinates, and even though the radial coordinate may change its causal structure, the polar coordinate remains spacelike. 

Thus, by considering geodesics with the spacelike tangent vectors and a special value of the Carter constant $C=4m^2$, we were able to generate a novel Schwarzschild-bumblebee black hole that bypasses an issue of the solution, presented in \cite{Casana2018}.

The next step is to try to involve the \textit{timelike} bumblebee fields. For this, we have to add a timelike component, namely, to consider non-zero $E$. In this case, the bumblebee field is given by
\begin{align}
     \mathbf{b}=&b\sqrt{1+\epsilon \xi b^2}\Big(-E\dd t+\sigma_r\sqrt{\dfrac{\epsilon}{f}\Big(1-\dfrac{\epsilon C^2 }{r^2}\Big)+\dfrac{E^2}{f^2}}\dd r+\sigma_x\sqrt{C}\dd\theta\Big).\label{68}
\end{align}

\begin{figure}
    \centering
    \includegraphics[width=1\linewidth]{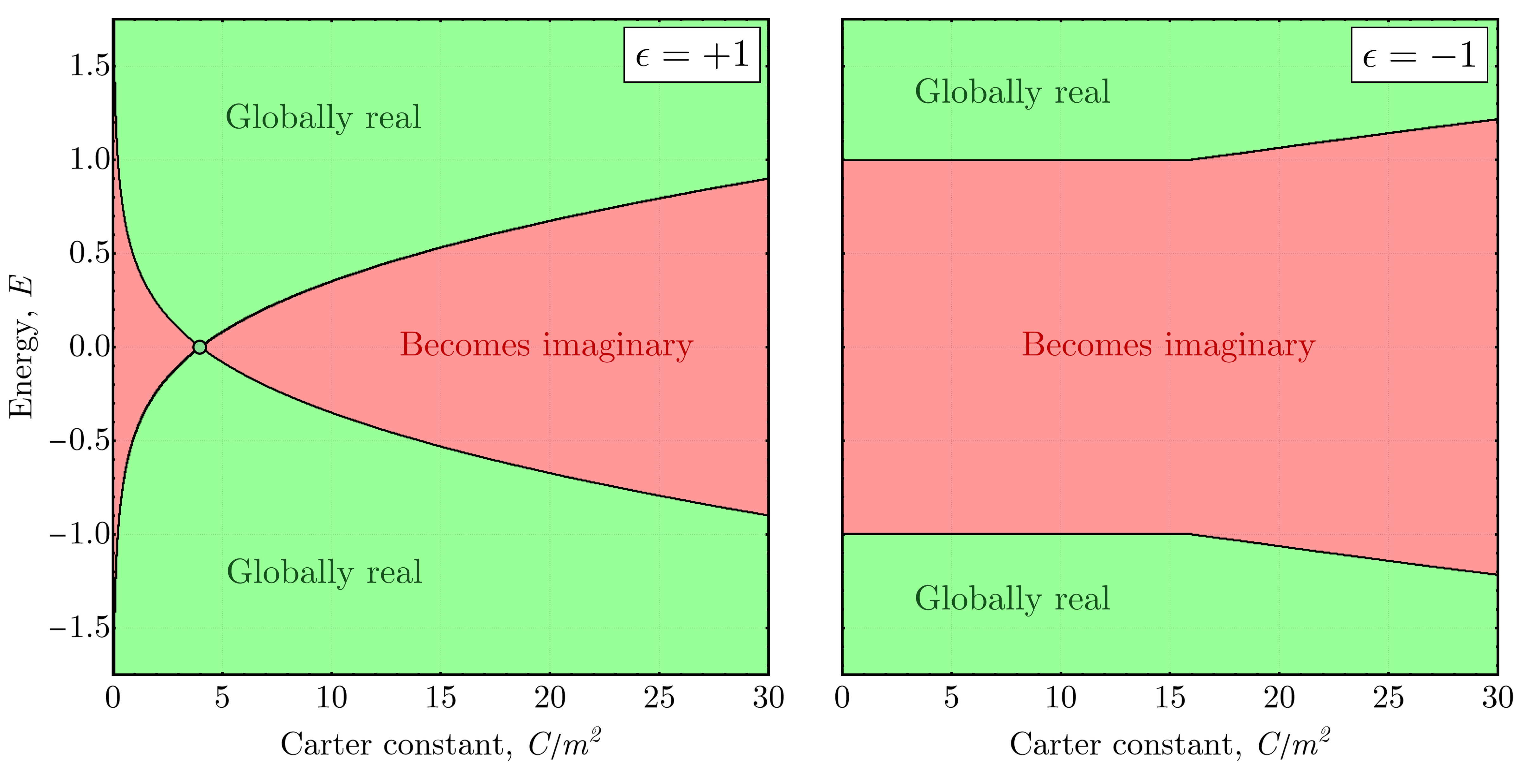}
    \caption{The ranges of energy $E$ and the Carter constant $C$, where the bumblebee field (\ref{68}) is either globally real (green ranges) or becomes imaginary at some range (red ranges). Black curves represent the border between such ranges. Left panel is the plot for the spacelike bumblebee field $\epsilon=+1$, while the right panel is the plot for the timelike bumblebee field $\epsilon=-1$. The green dot represents the special value $C=4m^2$ at which the bumblebee field becomes globally real even for $E=0$ (but only for the case $\epsilon=+1$).}
    \label{fig1}
\end{figure}

As we wish $\mathbf{b}$ to remain real, we have to require $C\geq 0$ and to choose such $C$ and $E$ that the expression in the square root remains positive. This is not a unique choice and allows for some range of the parameters $E$ and $C$. The corresponding ranges of energy $E$ and Carter constant $C$ are plotted on Fig. \ref{fig1}\footnote{Details of the algorithm, used to generate all the pictures, presented in this work, are described in the Supplementary Material.}. In principle, these conditions can be obtained analytically because deducing whether the expression in the square root of (\ref{68}) is non-negative or not leads to the investigation when the condition
\begin{align}
    (\epsilon r^2-C)(r-2m)+E^2 r^3\geq 0\label{Schw_real_cond}
\end{align}
is satisfied.

This analysis can be performed by the standard methods for the analysis of signs of polynomials, however, we do not wish to spend much time on the analytical approach and present only the final numerical solution on Fig. \ref{fig1}. What is notable in these pictures is that for the spacelike bumblebee field $\epsilon=+1$, the absolute value of energy is always limited from below, except for $C=4m^2$, which is exactly the value we have found during the analysis of the previous case. Also, the case of the timelike fields $\epsilon=-1$ becomes possible, and it can be easily seen that the energy $E$ has to be always $E^2\geq 1$. This follows from the obvious observation that for the energies $E^2<1$, a timelike geodesic observer cannot reach infinity.

The only analytical result that can be easily investigated is the case $C=0$. In this case, the condition (\ref{Schw_real_cond}) becomes $(\epsilon+E^2)r-2m\epsilon\geq 0$. For $\epsilon=+1$, there is no way this condition can be satisfied globally, while for $\epsilon=-1$ this appears to be possible for any $E^2\geq 1$. All these analytical results are reflected in Fig. \ref{fig1}.

The metric in this case is quite complicated, and we will not show it here explicitly, but it can be obtained using (\ref{Final_metr_and_bb}). Also, we wish to mention that the case with $C=0$ gives the solution obtained in \cite{Liu2025_2} (as can be seen after some transformation of coordinates, see Sec. 4.1 and eqs. (43a)-(43b) in \cite{Poulis2022}).

The cases when $L$ is non-zero however, suffer for the issue mentioned in \cite{Poulis2022}: the function $h'$ becomes imaginary near poles (where $P\to 0$). In principle, this has to be avoided as it does not give a real bumblebee field. However, as we noted during our previous discussion, the same issue also appears for standard bumblebee solutions \cite{Casana2018}, where this field becomes imaginary below the horizon, so the issue here is conceptually the same as the one for the widely used solution found in \cite{Casana2018}. The only significant difference is that the solution \cite{Casana2018} is real at the radial infinity, while the bumblebee field in the case with non-zero $L$ is imaginary near the poles \textit{for any $r$}. Also, notice that even in the most general case \eqref{Kerr-NUT_f}-\eqref{Kerr-NUT_h}, the $L\neq 0$ leads to imaginary $h'(x)$ near the poles $x=\pm 1$. Because of this, in all other subcases we will investigate only the case $L=0$.

\subsubsection{Kerr spacetime (no Taub-NUT parameter: $l=0$)}\label{Sec_Kerr}

Now, let us move to the case of Kerr spacetime. Even though the most general expression for the bumblebee field in this case was already obtained in \cite{Poulis2022}, it is worth considering it because in \cite{Poulis2022} authors did not investigate the physical properties of their solution in great detail. In addition, we will see here several properties that will be important for our further analysis when we add a cosmological constant or charges.

To apply our generating technique, we at first consider the $q,~l,~\Lambda\to 0$ limits in (\ref{Kerr-NUT_f})-(\ref{Kerr-NUT_h}):
\begin{align}
    f'(r)&=\sigma_r\sqrt{\dfrac{(\epsilon r^2-C)Q+(a L-E(a^2+r^2))^2}{Q^2}},\label{f_pr}\\
    h'(x)&=\sigma_x\sqrt{\dfrac{(\epsilon a^2 x^2+C)P-(L-a E(1-x^2))^2}{P^2}}\label{h_pr},
\end{align}
where metric functions $Q$ and $P$ become
\begin{align}
    Q=a^2-2m r+r^2,~~~~P=1-x^2.
\end{align}

Let us start with the case when $C=E=L=0$. In this case, the bumblebee field is given by
\begin{align}
    \mathbf{b}=b\sqrt{\epsilon}\sqrt{1+\epsilon \xi b^2}\Big(\sigma_r\dfrac{r}{\sqrt{Q}}\dd r+\sigma_xa\cos \theta \dd\theta\Big).\label{73}
\end{align}

This solution has nearly the same issue as in \cite{Casana2018}: below the horizon (where $Q<0$), the bumblebee field becomes complex. However, now, choosing another sign of $\epsilon$ does not cure this issue: here we have two terms ($dr$ and $d\theta$), and by changing the sign of $\epsilon$ we change the initially real $d\theta$ term to the imaginary one that we wish to avoid. 

The metric in this case is given by
\begin{align}
    ds^2=&-\dfrac{Q}{\rho^2}\big(\dd t-a(1-x^2)\dd\varphi\big)^2+\dfrac{P}{\rho^2}\big(a\dd t-(r^2+a^2)\dd\varphi\big)^2+\nonumber\\
    &\dfrac{(1+\epsilon\xi b^2)r^2+a^2 x^2}{Q}\dd r^2+\dfrac{r^2+(1+\epsilon \xi b^2)a^2x^2}{P}\dd x^2+2\epsilon \xi b^2\sigma_r\sigma_x \dfrac{a rx}{\sqrt{QP}}\dd r\dd x.\label{74}
\end{align}

In this case, as can be easily seen, the fact that the bumblebee field is not globally real also manifests itself in the fact that the additional non-diagonal term has the same issue, and it is well-defined only when $Q\geq 0$. Notice that for extremal black holes $a=m$ and naked singularities $a>m$, this condition is automatically satisfied, so this solution is well-defined in these cases, but for the non-extremal horizon, such an issue appears.

Thus, we notice that the attempt of the application of the easiest form of the generating technique (with $E=L=C=0$) to the case of Kerr spacetime gives not only an ill-defined bumblebee field (as in the case of \cite{Casana2018}), but also an ill-defined non-diagonal metric terms. Moreover, the algebraic type of the resulting metric is I. This shows the structural algebraic difference between the seed Kerr metric (that is of algebraic type D) and the solution \cite{Casana2018} (that is also of type D).

Now, let us discuss how the aforementioned issues can be bypassed. First of all, we notice that the introduction of the non-zero $C$ does not change the situation drastically when $E=L=0$. It is so because in this case  
\begin{align}
    \mathbf{b}=&b\sqrt{1+\epsilon \xi b^2}\Big(\sigma_r\sqrt{\dfrac{(\epsilon r^2-C)}{Q}}\dd r+\sigma_x\sqrt{\dfrac{(\epsilon a^2 x^2+C)}{P}}\dd x\Big).
\end{align}

Let us investigate the first square root in this expression. The corresponding expression is real when $\dfrac{(\epsilon r^2-C)}{Q}\geq 0$. If $\epsilon=+1$, the numerator becomes zero if $r=\pm \sqrt{C}$, while denominator always has roots at $r=r_{\pm}$, where $r_{\pm}=m\pm \sqrt{m^2-a^2}$. As each of the roots of the numerator and the denominator are of the 1-st order (unless the horizon is extremal), the corresponding expression changes its sign while crossing each root. As we have only one parameter $C$, there is no way to make this expression globally non-negative (the only positive root of the numerator cannot coincide with both roots of the denominator). Thus, it means that there always exists a range of $r$'s where the bumblebee field is complex unless $Q$ has a double root (extremal horizon) or does not have any root at all (naked singularity). In these two cases it is possible to choose such $C$ that the overall expression is real.

Now, let us discuss if there is a way to bypass this issue by adding a non-zero $E$. The only choice that allows us to cover the whole range of $r\in [0,\infty)$ and $x\in [-1,1]$ is the one with $C=a^2 E^2$. To see this, let us investigate (\ref{f_pr}) and (\ref{h_pr}). We require the expressions under the square roots to be globally non-negative. This also means that at any chosen points, they are also non-negative. Let us choose $r=0$ and $x=0$ as such points. In this case, the expression under the square root in (\ref{f_pr}) becomes $E^2-C/a^2$, while the expression under the square root in (\ref{h_pr}) $C-a^2E^2$. As both these expressions have to be $\geq 0$, the only possible choice is $C=a^2 E^2$. 

Now let us investigate under what conditions (\ref{f_pr}) and (\ref{h_pr}) are globally real. First of all, we notice that the reality of (\ref{h_pr}) under the conditions $C=a^2E^2$ and $L=0$ leads to
\begin{align}
    a^2x^2(1-x^2)(E^2+\epsilon)\geq 0.
\end{align}
This condition in the range $x\in [-1,1]$ is always satisfied for $\epsilon=+1$, while for $\epsilon=-1$ it will be satisfied only of $E^2>1$.

Now, let us investigate the reality condition of (\ref{f_pr}). It is real if the condition
\begin{align}
    E^2(r^3+a^2(2m+r))+\epsilon r(a^2-2m r+r^2)\geq 0.
\end{align}
holds.

\begin{figure}
    \centering
    \includegraphics[width=1\linewidth]{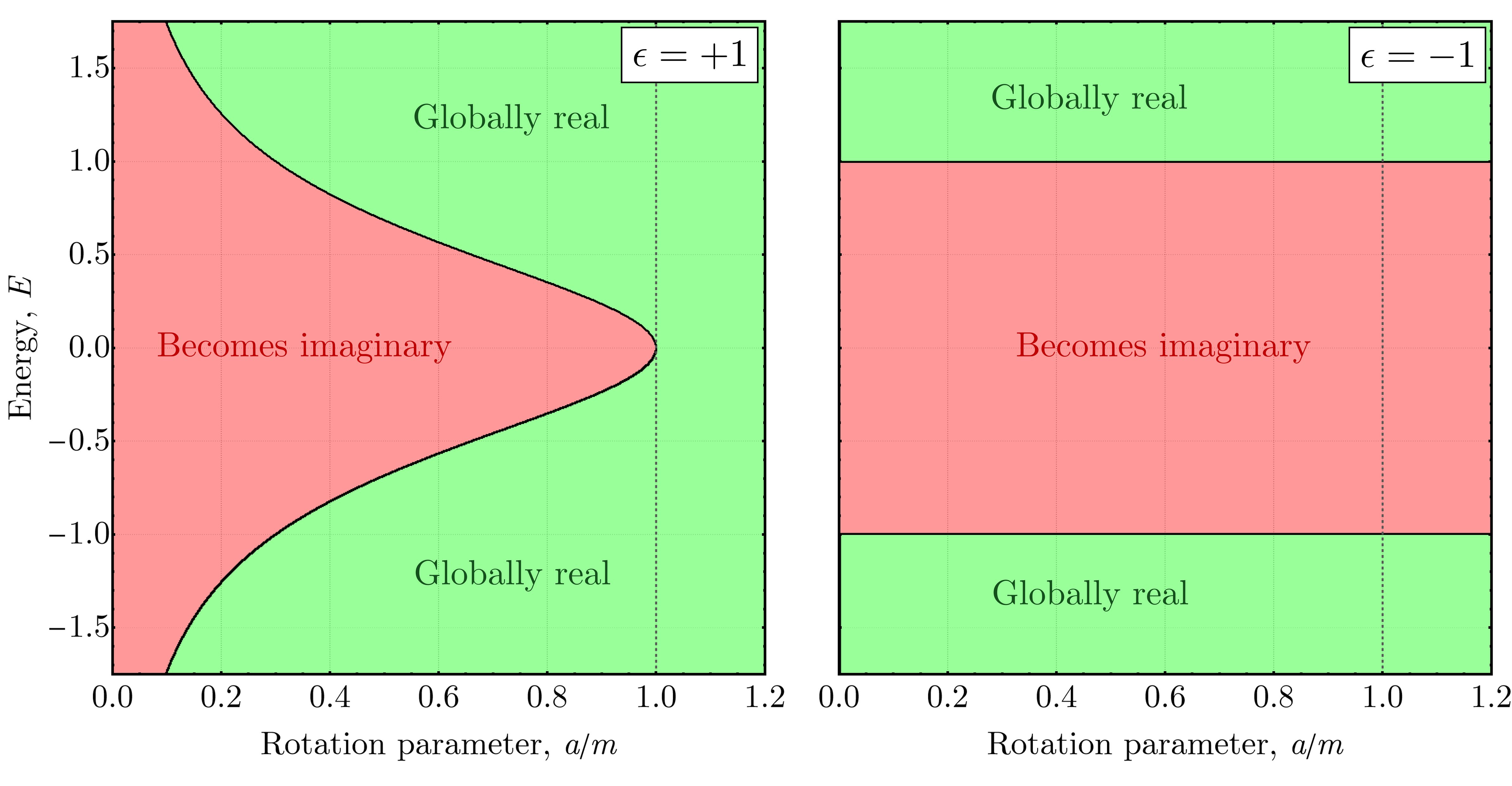}
    \caption{The ranges of energy $E$ and the rotation parameter $a/m$, where the bumblebee field (\ref{f_pr})-(\ref{h_pr}) is either globally real (green ranges) or becomes imaginary at some range (red ranges). Black curves represent the border between such ranges. The left panel is the plot for the spacelike bumblebee field $\epsilon=+1$, while the right panel is the plot for the timelike bumblebee field $\epsilon=-1$. The dashed line represents the value of $a/m=1$, where the black hole becomes extremal.}
    \label{fig2}
\end{figure}

This is also a qubic equation and can be analyzed by the standard methods. Let us focus on the result, presented in Fig. \ref{fig2}. 

From the analysis of these pictures, we clearly see that for the case $\epsilon=+1$ and for any range of $a<m$ (where the non-extremal black hole exists), $E^2$ is limited from below, while if $a\geq m$ (corresponding to the case of extremal Kerr spacetime and the naked singularity), energy $E$ can take any value. This is quite expected because in the case of a naked singularity $Q$ is positive everywhere, and thus the issues that were present previously because of the change of sign of $Q$ while crossing horizon do not appear for naked singularities. For the case $\epsilon=-1$, one only obtains the condition $E^2\geq 1$. The reason for implying this condition is clear: timelike particles with $E^2<1$ cannot reach infinity.

We do not consider the case with the non-zero $L$ because of the reason mentioned in Sec. \ref{Sec_static_vac}.

\subsubsection{Kerr--Taub-NUT spacetime}\label{Sec_Kerr-NUT}

The last vacuum spacetime we wish to investigate is the Kerr--Taub-NUT spacetime. In this case, (\ref{Kerr-NUT_f})-(\ref{Kerr-NUT_h}) become 
\begin{align}
    f'(r)&=\sqrt{\dfrac{\epsilon r^2-C}{Q}+\dfrac{(a L-E(r^2+(a+l)^2))^2}{Q^2}},\label{78}\\
    h'(x)&=\sqrt{\dfrac{\epsilon (ax+l)^2+C}{P}-\dfrac{(L-E(a(1-x^2)+2l(1-x)))^2}{P^2}},
\end{align}
with $Q=a^2-l^2-2m r+r^2$, $P=1-x^2$. Let us investigate the conditions that have to be imposed to have the globally real functions $f'$ and $h'$. Let us start with the $h'$. We expand the expression under the square root near $x=1$. Explicit expansion gives (here we take $L=0$ because of the reason mentioned at the end of Sec. \ref{Sec_static_vac}):
\begin{align}
    (h'(x))^2=\dfrac{C+\epsilon (a+l)^2}{2(1-x)}+O(1).
\end{align}

The divergent term will be non-negative if
\begin{align}
    C\geq -\epsilon (a+l)^2\label{C_cond_NUT}.
\end{align}

The next step is to investigate the behaviour near $x=-1$. Conducting an expansion, one obtains
\begin{align}
    (h'(x))^2=-\dfrac{4E^2l^2}{(1+x)^2}+O\big((1+x)^{-1}\big).
\end{align}

The dominant term here is always negative unless $E=0$. If it is so, then the $h'(x)^2$ becomes:
\begin{align}
    h'(x)^2=\dfrac{\epsilon (ax+l)^2+C}{P}.
\end{align}

In the range $x\in [-1,1]$, it is always positive if the condition (\ref{C_cond_NUT}) is satisfied. 

Now let us move to the radial function. Under the conditions $L=0,~E=0$, \eqref{78} becomes:
\begin{align}
    f'(r)^2=\dfrac{\epsilon r^2-C}{Q}.\label{84}
\end{align}

The numerator of this expression has two roots: namely the $\pm \sqrt{C/\epsilon}$. Only one of them is positive (if these roots are real). However, the denominator has two roots (unless BH is extremal): at the inner $r_-$ and outer horizons $r_+$:
\begin{align}
    r_{\pm}=m\pm \sqrt{m^2+l^2-a^2}.
\end{align}

All the roots of the numerator and denominator are of the 1-st order, and $f'(r)^2$ changes sign when crossing each of them. By adjusting some special value of $C$ we are only able to cancel one root of the denominator, but not both. This means that for \textit{non-extremal Kerr--Taub-NUT spacetimes there is no way to choose such $E,~L$ and $C$ that the bumblebee field is globally real.} The reason for such a result is the presence of the twisting string, attached to the south pole of the Kerr--Taub-NUT black hole, which makes the south pole non-regular (for the details see \cite{Podolsk2021,Podolsk2023,Ovcharenko2025_2,Ovcharenko2024}). 

In the case of extremal BH and for a naked singularity, the situation is somehow different. In both these cases, $Q$ is globally non-negative. Then the denominator in (\ref{84}) is always non-negative, and we have to require the numerator to be non-negative. This can be achieved only for $\epsilon=+1$ and $C\geq 0$ (then the condition \eqref{C_cond_NUT} is automatically satisfied).

The last thing we wish to mention about this solution is that in the $a=0$, $E=L=0$, $\epsilon=+1$, $C=-l^2$ limits, one obtains the solution with
\begin{align}
    \mathbf{b}=b\sqrt{1+\xi b^2}\,\dfrac{\sqrt{r^2+l^2}}{\sqrt{Q}}\,\dd r,
\end{align}
and the metric is given by
\begin{align}
    ds^2=-\dfrac{Q}{r^2+l^2}(\dd t-2l(1-\cos\theta)\dd\varphi)^2+(1+\xi b^2)\dfrac{r^2+l^2}{Q}\dd r^2+(r^2+l^2)(d\theta^2+\sin^2\theta \dd\varphi^2),
\end{align}
with $Q=r^2-2m r-l^2$. This is exactly the solution found in \cite{Chen2025} (that can be obtained after trivial transformation of coordinates and trivial reparametrizations). However, this solution still has the same issue as many of those discussed previously: the bumblebee field becomes complex when crossing the horizon. 

\subsection{Spacetimes with a cosmological constant (no charge: $q=0$)}

Now, let us move to the discussion of the cases with a cosmological constant. The metric is given by (\ref{ds2_Kerr-NUT}), the metric functions (\ref{Q_gen})-(\ref{P_gen}) are
\begin{align}
    Q&=a^2-l^2-2m r+r^2-\dfrac{\Lambda}{3}r^2\Big(r^2+a^2+3 l^2\Big),\\
    P&=(1-x^2)\Big(1+2\dfrac{\Lambda }{3}l(l+a x)+\dfrac{\Lambda}{3}(l+a x)^2\Big).
\end{align}

Solution to the Hamilton--Jacobi equation in this case is given by (\ref{rho_gen}) with (\ref{Kerr-NUT_f})-(\ref{Kerr-NUT_h}), but the metric functions $Q$ and $P$ are different. Let us investigate the conditions under which the bumblebee field is real. First of all, we would like to mention that, as in the Kerr--Taub-NUT case, if the non-zero Taub-NUT parameter $l$ is present, then there is no way to choose such parameters $E,~L$ and $C$ to make the bumblebee field globally real if the inner and outer horizons are present (proof is analogous to the one given in Sec \ref{Sec_Kerr-NUT}).  

Because of these reasons, we will focus our attention only on the cases with zero Taub-NUT parameter $l=0$ and try to understand what limitations on $E,~C$ are present in this case. Also, we wish to mention that for the case with the zero Taub-NUT parameter $l=0$, we still have to choose $L=0$, because otherwise (\ref{Kerr-NUT_h}) becomes imaginary near the poles $x=\pm 1$. Thus, in our further analysis we will assume that $L=0$.

\subsubsection{Static case (no twist: $a=0,~l=0$)}\label{Sec_stat_ads}

Let us start with the static case and with $E=L=C=0$. In fact, this case was already considered and it gives rise to the bumblebee field (\ref{Schw_bb}) and the metric (\ref{casana_metr}), however, here $f=1-\dfrac{2m}{r}-\dfrac{\Lambda}{3}r^2$. All the important issues related to the reality of the bumblebee field are still present in this case, so to avoid them, one may try adding additional parameters.

The only parameters we can add are $E$ and $C$. The bumblebee field in this case is given by \eqref{68} with $f=1-\dfrac{2m}{r}-\dfrac{\Lambda}{3}r^2$. Reality of the corresponding square root in (\ref{68}) is assured by the inequality:
\begin{align}
    (\epsilon r^2-C)\Big(r-2m-\dfrac{\Lambda}{3}r^3\Big)+E^2 r^3\geq 0.\label{lambda_ineq}
\end{align}

\begin{figure}
    \centering
    \includegraphics[width=1\linewidth]{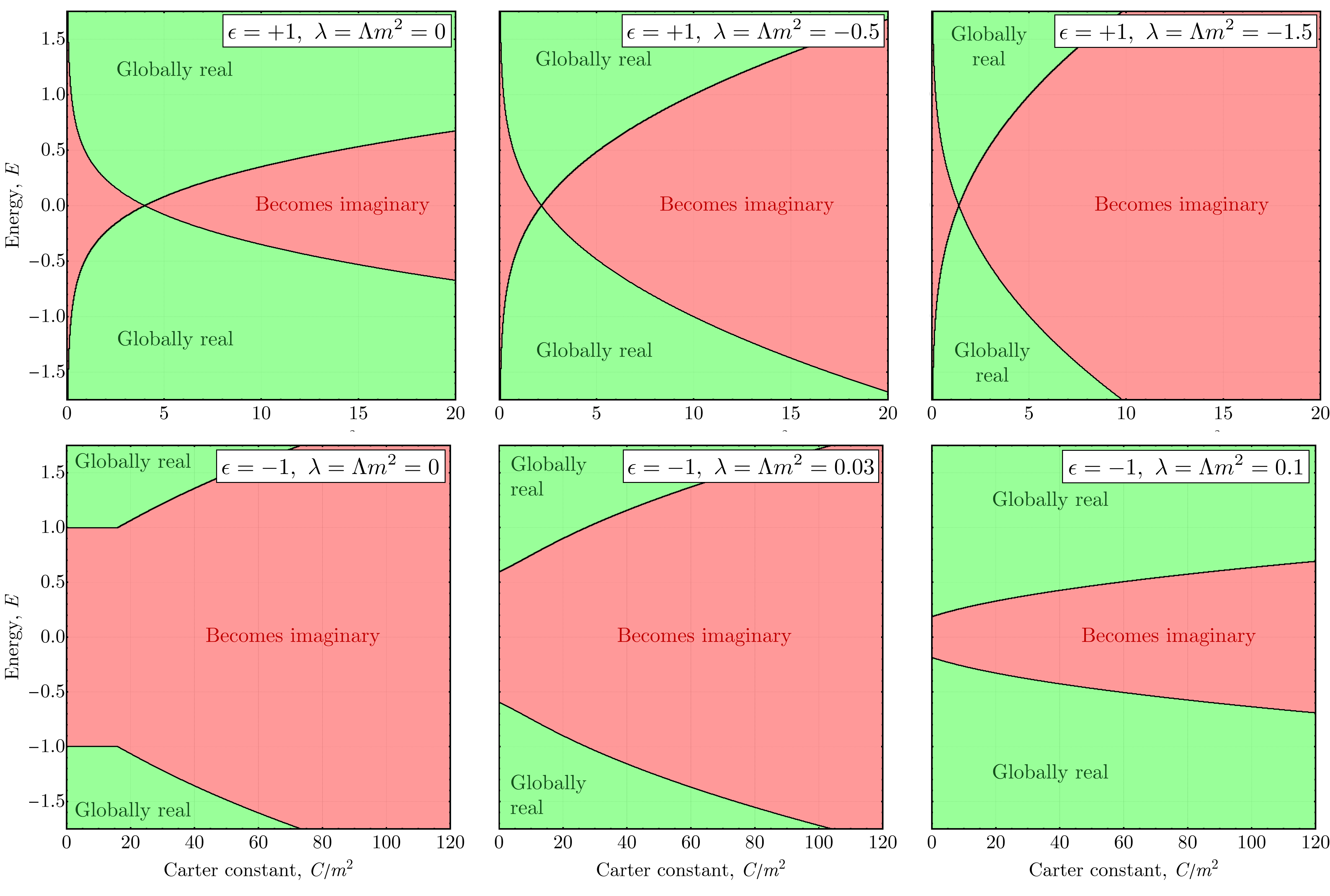}
    \caption{The ranges of energy $E$ and the Carter constant $C$, where the bumblebee field (\ref{68}) for the Schwarzschild--(A)dS black hole with $f=1-\dfrac{2m}{r}-\dfrac{\Lambda}{3}r^2$ is either globally real (green ranges) or becomes imaginary at some range (red ranges). Various plots represent various values of the cosmological constant for spacelike $\epsilon=+1$ and timelike $\epsilon=-1$ bumblebee fields.}
    \label{fig3}
\end{figure}

As the corresponding expression on the LHS is of the 5-th order, its full analytical investigation is extremely complicated. However, we can conduct some analysis anyway by analyzing the asymptotics. For example, we see that at large $r$'s the LHS of (\ref{lambda_ineq}) behaves as $-\epsilon\Lambda r^5/3$. Thus for $\epsilon=+1$ the dominant term will be positive only if $\Lambda$ is negative, while if $\epsilon=-1$, it will be positive only if $\Lambda$ is positive. This result is, in fact, not so surprising because it is known that the timelike particles ($\epsilon=-1$) cannot reach the spatial infinity in the AdS space \cite{Hackmann2008,Hackmann2008_2,Hackmann2010}, while for the spacelike ($\epsilon=+1$) the same holds in dS space. That is why we will investigate only timelike particles in dS space and spacelike in AdS. The corresponding result of numerical analysis is presented in Fig. \ref{fig3}. First of all, we notice that for the timelike fields ($\epsilon=-1$), the increase of the cosmological constant leads to the widening of the ranges of the parameters at which the bumblebee field remains globally real. This happens because of the repulsive nature of the de Sitter space, and because of this, the energy required to reach infinity is lower. 

However, for the spacelike fields ($\epsilon=+1$), the situation is somewhat different, and increasing the absolute value of the cosmological constant leads to the squeezing of the ranges where the bumblebee field is globally real. The reason for this is that, probably, as the negative cosmological constant acts repulsively for the spacelike trajectories (see the discussion above), it is more complicated for spacelike fields to reach the internal domains of a black hole. Also we wish to notice that even in this case it is possible to have a globally real bumblebee field for the zero energy (but for the smaller value of the Carter constant). This happens because for the negative values of cosmological constant, the function $Q$ has only one positive root that becomes smaller with the increase of the absolute value of $\Lambda$, and the roots of the numerator and denominator of (\ref{68}) may coincide for a smaller value of $C$.

\subsubsection{Kerr--(A)dS (no Taub-NUT parameter: $l=0$)}

Now let us investigate the Kerr--(A)dS. The metric functions are given by
\begin{align}
    Q=&a^2-2mr+r^2-\dfrac{\Lambda}{3}r^2(r^2+a^2),\label{Q_Kerr-AdS}\\
    P=&(1-x^2)\Big(1+\dfrac{\Lambda}{3}a^2x^2\Big).\label{P_Kerr-AdS}
\end{align}

For the case $E=C=0$, the bumblebee field is given by (\ref{73}) and the metric is given by (\ref{74}), but the metric functions $Q$ and $P$ are now given by (\ref{Q_Kerr-AdS})-(\ref{P_Kerr-AdS}) (under the obvious assumption $L=0$). However, it still has the same issue that the bumblebee field becomes imaginary between the inner and outer horizons. To avoid this, we add both parameters $E$ and $C$. In this case, the radial and angular terms of the bumblebee field are still given by (\ref{f_pr})-(\ref{h_pr}). Reality of these expressions will be guaranteed if the numerators under the square roots are positive. As in the Sec. \ref{Sec_Kerr}, by considering the points $r=0$ and $x=0$, the only way of having real bumblebee field is to require $C=a^2E^2.$

Even with this simplifying condition, the exact analysis of the reality conditions is still quite complicated. The only thing we wish to mention here is that, as in the static case, the behaviour of (\ref{f_pr}) at large distances is defined by the relative sign of $\epsilon$ and $\Lambda$. Namely, if $\epsilon=+1$, then $\Lambda$ has to be negative to have real bumblebee field at infinity, while if $\epsilon=-1$, then $\Lambda$ has to be positive. Physical reasons for this were already discussed in Sec. \ref{Sec_stat_ads}.

\begin{figure}
    \centering
    \includegraphics[width=1\linewidth]{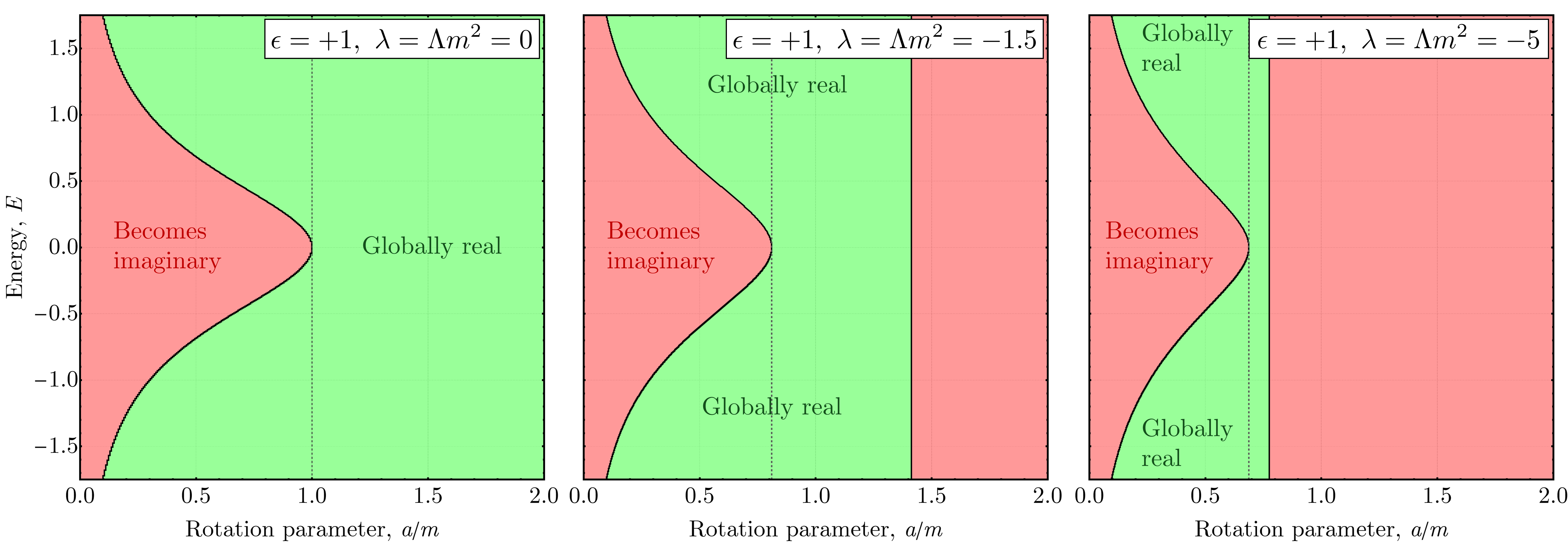}
    \caption{The ranges of energy $E$ and the rotation parameter $a/m$, where the bumblebee field for Kerr--(A)dS spacetime is either globally real (green ranges) or becomes imaginary (red ranges) for different $\lambda=\Lambda m^2$. Black curves represent the border between such ranges. Plots represent only the cases with $\epsilon=+1$, because for the case $\epsilon=-1$ the corresponding pictures are the same as for the case without cosmological constant, see Fig. \ref{fig2}. The dashed line represents the value of $a/m$, where the black hole becomes extremal.}
    \label{fig4}
\end{figure}

The numerical results for the ranges of parameters where the bumblebee field is real are given in Fig. \ref{fig4}. First of all, we wish to mention that for the timelike ($\epsilon=-1$) field, we did not plot the ranges of reality because the corresponding pictures for the positive $\Lambda$ are almost the same as for the case without the cosmological constant, see Fig. \ref{fig2}. For the negative $\Lambda$, as was discussed above, there is no way of having a globally real timelike bumblebee field.  However, for the spacelike ($\epsilon=+1$) field, there appears to be a significant change. We notice that the increase of the absolute value of the cosmological constant shifts the value of the rotation parameter $a$ where the horizon becomes extremal to lower values, but the shape of the edge remains more or less the same. However, what is more interesting is the appearance of the limiting value $a/m$, above which the bumblebee field becomes imaginary for naked singularities. The appearance of such regions may seem strange; however, it is related to the appearance of additional roots (in addition to $x=\pm 1$) of the function $P$ (\ref{P_Kerr-AdS}) for the negative cosmological constant. The exact position of this boundary is given by $a/m=\sqrt{-\dfrac{3}{\Lambda m^2}}$.

\subsection{Charged spacetimes}

\subsubsection{Reissner--Nordstr\"{o}m spacetime (no twist $a=0,~l=0$ and $\Lambda=0$ )}\label{Sec_RN}

The first charged spacetime we wish to investigate is the Reissner--Nordstr\"{o}m one. Basically, all the expressions for the bumblebee field and the metric are the same as for the case of the Schwarzschild spacetime (see Sec. \ref{Sec_static_vac}), but with the function $f$ given by $f=1-\dfrac{2m}{r}+\dfrac{q^2}{r^2}$. For the case $E=C=0$, we obtain the solution in the form (\ref{Schw_bb}) and (\ref{casana_metr}). What is interesting, the corresponding spacetime and the bumblebee field are the same as the ones presented in \cite{Liu2025,Li2025_2} after a rescaling of the charge parameter. This fact is quite surprising, as these solutions were found for different theories. These solutions suffer from the same issue as the solution \cite{Casana2018}: the bumblebee field becomes imaginary below the horizon. 

There naturally appears a question whether it is possible to have a globally real bumblebee field by a specific choice of $E$ and $C$. By inspecting (\ref{68}), we see that it will be globally real if the condition
\begin{align}
    (r^2-2m r+q^2)(\epsilon r^2-C)+E^2r^4\geq 0\label{RN_real_cond}
\end{align}
is satisfied

First of all, we notice that at $r=0$ the LHS of this condition is $-q^2 C$. As $C\geq 0$ (this is required to have real $d\theta$ term in (\ref{68})), we see that the condition (\ref{RN_real_cond}) will be satisfied at $r=0$ only if either $q=0$ (already considered in Sec. \ref{Sec_static_vac}) or if $C=0$. In the second case, the condition (\ref{RN_real_cond}) simplifies to 
\begin{align}
    (\epsilon+E^2)r^2-2m \epsilon r+\epsilon q^2\geq 0.\label{RN_real_cond_2}
\end{align}

We notice that at $r=0$ the LHS of this expression is $\epsilon q^2$, and it satisfies (\ref{RN_real_cond_2}) only if $\epsilon=+1$. If it is so, then by standard analysis of quadratic polynomials, one obtains that the condition (\ref{RN_real_cond_2}) is satisfied for $r\in [0,\infty)$ if 
\begin{align}
    E^2\geq\dfrac{m^2-q^2}{q^2}\label{RN_cond}.
\end{align}

This result is quite interesting as it is qualitatively different from the one obtained for the vacuum case. The case of the positive $C$ does not allow for a globally real bumblebee field, while for $C=0$ there appears a range of energies where this appears to be possible. In the vacuum case the situation is completely different as only for $C>0$ there appears such a possibility while for $C=0$ there is no way to have a globally real bumblebee field. Also, the RN solution fully forbids $\epsilon=-1$ case.

\subsubsection{Reissner--Nordstr\"{o}m--(A)dS (no twist $a=0,~l=0$)}

The next case we wish to discuss is the Reissner--Nordstr\"{o}m--(A)dS spacetime. Basically, all the expressions for the bumblebee field and the metric are the same as for the case of Reissner--Nordstr\"{o}m spacetime (see Sec. \ref{Sec_RN}), but with the function $f$ given by $f=1-\dfrac{2m}{r}+\dfrac{q^2}{r^2}-\dfrac{\Lambda}{3}r^2$. Thus, without additional details, we mention that by conducting the same analysis, one can easily deduce that the only possible case is the case with $C=0$ and $\epsilon=+1$. Under these assumptions, the analog of the condition (\ref{RN_real_cond_2}) becomes
\begin{align}
   -\epsilon \dfrac{\Lambda}{3}r^4+ (\epsilon+E^2)r^2-2m \epsilon r+\epsilon q^2\geq 0.\label{RN-AdS_cond}
\end{align}

\begin{figure}
    \centering
    \includegraphics[width=1\linewidth]{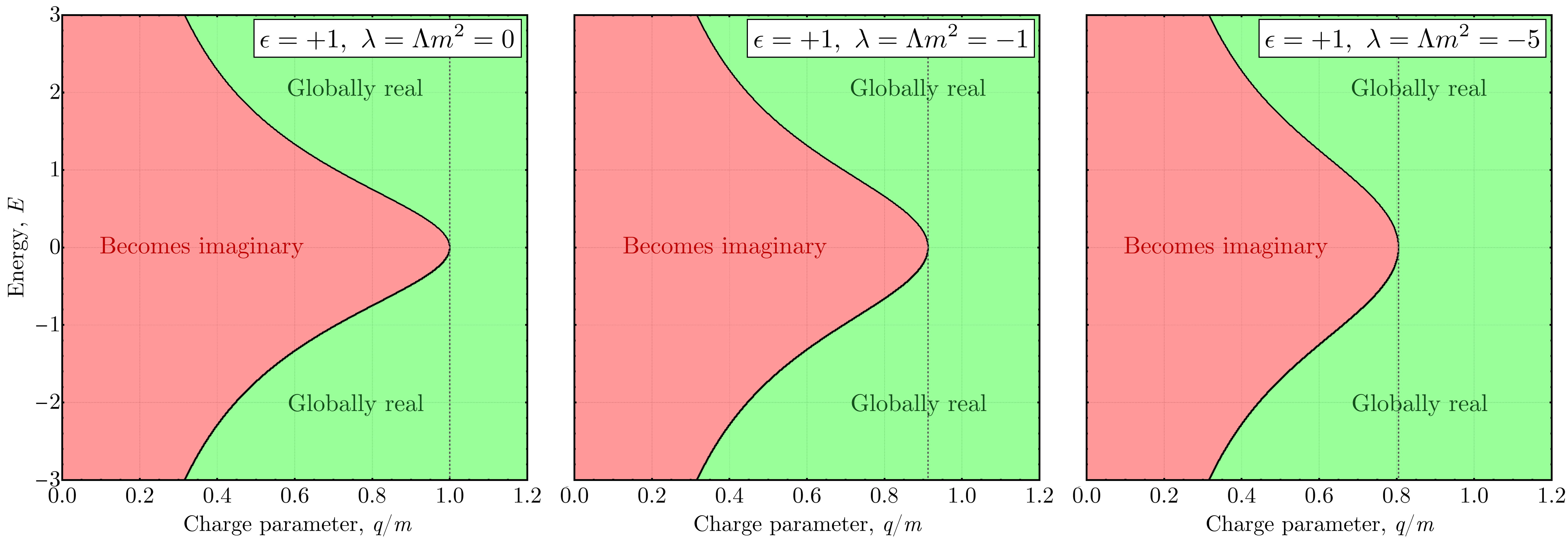}
    \caption{The ranges of energy $E$ and the charge parameter $q/m$, where the bumblebee field for the Reissner--Nordstr\"{o}m--(A)dS spacetime is either globally real (green ranges) or becomes imaginary (red ranges) for different $\lambda=\Lambda m^2$. Black curves represent the border between such ranges. The dashed line represents the value of $q/m$, where the black hole becomes extremal.}
    \label{fig5}
\end{figure}

As in Sec. \ref{Sec_RN}, we are able to consider only spacelike fields ($\epsilon=+1$). From the large $r$ asymptotics, it is clear that the corresponding condition (\ref{RN-AdS_cond}) will be satisfied only for $\Lambda\leq 0$. Generally, this condition is quite complicated to analyze analytically, and we show here only the result of the numerical analysis, presented in Fig. \ref{fig5}.

From Fig. \ref{fig5} one can clearly see that the only thing that appears because of the bumblebee field is the widening of the ranges at which the bumblebee field is real. Otherwise, no significant qualitative change is observed.

\subsubsection{Kerr--Newman spacetime (no Taub-NUT parameter $l=0$ and $\Lambda=0$)}

Now, we move to the case of the Kerr--Newman black holes, corresponding to $\Lambda,~l\to 0$ in the general expressions. In this case functions $Q$ and $P$ are given by
\begin{align}
    Q=&a^2+q^2-2mr+r^2,\label{Q_Kerr_Newman}\\
    P=&(1-x^2).\label{P_Kerr-Newman}
\end{align}

The bumblebee field is given by (\ref{f_pr})-(\ref{h_pr}), and the metric is given by (\ref{metr_genn}). Let us investigate whether the bumblebee field can be chosen globally real or not. For this, we do the same analysis we used in Sec. \ref{Sec_Kerr}, namely to take a point at which $r=0$ and $x=0$. For this point, the expression under the square root in $dr$ term becomes
\begin{align}
    \dfrac{a^4E^2-C(a^2+q^2)}{(a^2+q^2)^2},
\end{align}
(here we, as in all previous cases assume $L=0$ because of the reasons explained in Sec. \ref{Sec_static_vac}).

The expression under the square root in $dx$ term becomes
\begin{align}
    C-a^2E^2.
\end{align}

The second expression is non-negative if $C\geq a^2E^2$, while the first is non-negative if $C\leq a^2E^2 \dfrac{a^2}{a^2+q^2}$. It can be easily seen that these two conditions are incompatible unless $q=0$ or $a=0$ or $E=0$. The first two cases correspond to the already considered cases of Kerr and Reissner--Nordstr\"{o}m spacetime, while for the case $E=0$ one has that the only possible value of the Carter constant is $C=0$ and that the $f'$ in (\ref{Kerr-NUT_f}) becomes:
\begin{align}
    f'(r)=\sigma_r \sqrt{\dfrac{\epsilon r^2-C}{Q}}.
\end{align}

As in the case of Kerr spacetime (see Sec. \ref{Sec_Kerr}), if $Q$ has roots of the first order (non-extremal horizon), then this expression becomes imaginary below the horizon. And only when $Q$ has a root of the second order (extremal black hole) or if $Q$ does not have any roots (naked singularity), this expression is globally real (but only for the case $\epsilon=+1$).

This analysis gives us quite interesting results. \textit{For the generic non-extremal Kerr--Newman black hole, there is no non-dynamical globally real bumblebee field, given by its VEV}, while for the pure Kerr, such fields exist. This does not mean that such bumblebee field cannot exist at all; most probably, this means that one of the used assumptions (that the bumblebee field is given by its VEV $B_{\mu}=b_{\mu}$ or that it is not dynamical $b_{\mu\nu}=0$) fails. However, the investigation of these possibilities exceeds the scope of the current work.

Anyway, previously, this was not a great issue as the researchers were mainly interested in the behaviour of the bumblebee field outside the horizon where this solution still nicely works. In this case, the bumblebee field and the metric tensors are given by (\ref{73}) and (\ref{74}) with the metric functions $Q$ and $P$ given by (\ref{Q_Kerr_Newman}) and (\ref{P_Kerr-Newman}). 

Here we also wish to mention that considering more generic spacetimes (such as Kerr--Newman-(A)dS) does not give anything new because considering such generalization does not allow to get a globally real bumblebee field.

At the end, we wish to discuss what happens in the case of other spacetimes that are not part of the Kerr--Newman--Taub-NUT--(A)dS family. As was mentioned in Sec. \ref{Sec_gen_metr}, consideration of this class of spacetime is motivated by the fact that it is the most general spacetime within Einstein--Maxwell gravity with $\Lambda$ where the Hamilton--Jacobi equation for the spacelike or timelike geodesics is separable. However, this does not mean that this is the only spacetime one can consider to generate novel solutions. The technical limitations that prevent doing so are the inability of finding analytical solutions to the Hamilton--Jacobi equation for more general spacetimes. However, numerical methods are still available, and by the generating technique we have analzyed in this work, a bumblebee solution can be easily found, knowing the corresponding electrovacuum solution. Examples of such spacetimes that may be interesting for such an investigation are the astrophysically relevant ones, describing, for example, the recently found class of black holes, immersed into a uniform Bertotti--Robinson electromagnetic field \cite{Podolsk2025,Ovcharenko2025_3}.

\section{Conclusions}\label{Sec_concl}

In this work, we showed in the Einstein-bumblebee theory that if the bumblebee field is given by its vacuum expectation value $B_{\mu}=b_{\mu}$ and it is not dynamical $b_{\mu\nu}=\partial_{\mu}b_{\nu}-\partial_{\nu}b_{\mu}=0$, then these conditions \textit{uniquely} define the generating technique that allows to find solutions to the bumblebee gravity from the background metric by the transformation (\ref{Final_metr_and_bb}) that defines the metric and the bumblebee field. The bumblebee field in (\ref{Final_metr_and_bb}) is proportional to the tangent vector to the (spacelike or timelike) geodesic curve, and it can be found by knowing the solution to the Hamilton--Jacobi equation (\ref{HJ_eq}) on the background spacetime. The parameter $\epsilon$ here is used to represent whether the bumblebee field is spacelike $\epsilon=+1$, or timelike $\epsilon=-1$. The modification to the metric was suggested previously in \cite{Poulis2022}, while in our work, we showed that it is unique and inspect the relation between the geodesic curves \textit{on the ``seed'' spacetime} $\tilde{g}_{\mu\nu}$ and the bumblebee field. In addition, we proved that this generating technique can also be used in the case of a non-zero cosmological constant and in the presence of an electromagnetic field. If the electromagnetic field is present, then the Faraday tensor is not simply the Faraday tensor of the background solution but it is rather given by \eqref{Faraday_rel}. This analysis provides a powerful method of finding novel spacetimes within the bumblebee gravity without the need to solve the field equations. As was also mentioned in \cite{Poulis2022}, this generating technique reveals the deep physical structure of the bumblebee theory: the metric is modified by the additional term, related to a specific direction in the spacetime, leading to the \textit{explicit} Lorentz-violation of the resulting metric.

Then we applied this technique to the most general seed spacetime where the separability of the Hamilton--Jacobi equation is possible: the Kerr--Newman--Taub-NUT--(A)dS spacetime. In the analysis on this spacetime, we have shown that while applying this method, one does not generate the unique bumblebee solution, but rather generates a class of solutions related to various geodesics in the background spacetime. These are characterized by the energy $E$, the angular momentum $L$, and the Carter constant $C$ of the geodesic to which one associates the bumblebee field. 

However, even though a large class of geodesics was generated by this method, not all of them represent a globally real bumblebee field. For example, a non-zero angular momentum $L$ inevitably leads to imaginary bumblebee field near the poles $\theta=0,~\pi$, thus the real $b_{\mu}$ can be obtained only if $L=0$. In addition, we have shown that even for the simplest solution \cite{Casana2018}, $b_{\mu}$ becomes imaginary below the horizon, which is related to the change of the causal structure of the radial direction below the horizon. Nevertheless, we were able to show that this issue can be bypassed if one introduces a non-zero $E$ and $C$, as presented in Fig. \ref{fig1}. Analogous analysis was done also for the cases of Kerr, Schwarzschild--(A)dS, Kerr--(A)dS and Reissner--Nordstr\"{o}m--(A)dS spacetimes. The corresponding ranges of the parameters where the bumblebee field is globally real are shown in Figs. \ref{fig2}--\ref{fig5}.

In all other cases, we were able to show that it is impossible to choose some specific parameters that make the bumblebee field globally real. Remarkably, this holds for the Kerr--Newman spacetime. This is most probably related to the fact that the non-trivial charge and the backreaction of the electromagnetic field on the metric lead to the fact that the geodesic curves (globally real in the polar coordinate $\theta$) cannot reach the singularity. This means that, most probably, for such solutions some of the employed assumptions fail, namely whether the bumblebee field is not given by its vacuum expectation value $B_{\mu}\neq b_{\mu}$ or it is dynamical $d\mathbf{b}\neq 0$. Thus, we hope that further analysis of the cases when these assumptions do not hold will bring more understanding to the description of such more complicated solutions.

\section{Supplementary material}

Detailed derivations and the code used to generate plots are contained in the supplementary Wolfram Mathematica file, see \cite{supp_mat}.  

\section{Acknowledgments}

This work was supported by the Czech Science Foundation Grant No. GA\v{C}R 26-22381S and by the Charles University Grant No. GAUK 260325. The author thanks Ji\v{r}\'{i} Podolsk\'{y} for valuable discussions and comments.

\bibliographystyle{apsrev4-1}
\bibliography{references}{}

\end{document}